\documentstyle[epsfig,mncite]{mn}
\begin{document}
\newcommand\msun   {M$_{\odot}$}
\LARGE
\normalsize
\title{Low and high frequency variability as a function of 
spectral properties in the bright X--ray binary GX~5--1}

\author[P. G. Jonker et al.]
{P. G. Jonker$^{1,2}$\thanks{email : peterj@ast.cam.ac.uk, present
address Institute of Astronomy, Cambridge},
M. van der Klis$^1$,
J. Homan$^1$,
M. M\'endez$^3$,
W.H.G. Lewin$^4$, \cr
R. Wijnands$^{4,5}$,
W. Zhang$^6$\\
$^1$Astronomical Institute ``Anton Pannekoek'',
University of Amsterdam, Kruislaan 403, 1098 SJ Amsterdam\\
$^2$Marie Curie Fellow\\
$^3$Space Research Organization Netherlands, Sorbonnelaan
2, 3584 CA Utrecht, The Netherlands\\
$^4$Department of Physics and Center for Space Research,
Massachusetts Institute of Technology, Cambridge, MA 02138\\
$^5$Chandra Fellow\\
$^6$Laboratory for High Energy Astrophysics, Goddard
Space Flight Center, Greenbelt, MD 20771\\
}
\maketitle

\begin{abstract}
\noindent
We report on a detailed analysis of data obtained over nearly four
years with the {\it Rossi X--ray Timing Explorer} of the Z source
GX~5--1. From a spectral analysis using a hardness--intensity diagram
it was found that the source traced out the typical Z--shaped
pattern. The study of the power spectral properties showed that when
the source moved on the Horizontal Branch towards the Normal Branch
the fractional rms amplitudes and timescales of all variability
decreased, while their FWHMs increased. The frequency separation of
the two kHz QPO peaks decreased from $344\pm12$ Hz to $232\pm13$ Hz
while the frequency of the lower and upper kHz QPO increased from
172$\pm$10 Hz to 608$\pm6$ Hz and 516$\pm$10 Hz to 840$\pm$12 Hz,
respectively. At low frequencies, besides the horizontal branch
oscillation (HBO) and its second harmonic, two additional broad
Lorentzian components were needed to obtain acceptable fits. These
broad Lorentzians have Q--values of $\sim$1--2 and have frequencies
0.5 and 1.5 times the HBO frequency. When interpreted as being related
to the HBO, they seem to favor disc models for the HBO over the
magnetic beat--frequency model. The frequency of the Normal Branch
Oscillations changed slightly and non--monotonically while on the
Normal Branch between $\sim$6 Hz at both ends and 5.25$\pm$0.05 Hz
near the middle of the branch. It evolved into a flat--topped noise
component on the Flaring Branch. We compared the timing properties of
the some of the Z sources. We also compare the timing properties and
colour--colour diagrams (CDs) of GX~5--1 with those of the back hole
candidate XTE~J1550--564 and the atoll source 4U~1608--52. The CDs are
strikingly similar when a colour scheme is used that is commonly
employed in back hole studies. However, this may be a degeneracy as
the CDs turn out to be more complicated when colours common in neutron
star studies are employed.  Apart from some remarkable similarities
between the CD of XTE~J1550--564 and that of 4U~1608--52, several
differences can be seen between these CDs and that of
GX~5--1. Conclusions on spectral states or properties based solely on
the use of CDs using the ``black hole scheme'' should be regarded with
caution.

\end{abstract}


\section{Introduction}
\label{intro}
\noindent
Low--mass X--ray binaries (LMXBs) are systems where the compact
object, either a neutron star or a black hole, accretes matter from a
companion with a mass of less than 1 $M_{\odot}$. The neutron--star
LMXB systems are subdivided on the basis of the pattern they trace out
in an X--ray colour--colour or hardness--intensity diagram (CD or HID,
respectively) in atoll and Z sources \cite{hava1989}. The X--ray flux
measurements combined with the knowledge of the distance have shown
that Z sources have a high luminosity ($\sim$L$_{Edd}$) and accrete at
a rate close to the Eddington accretion rate, whereas atoll sources
have typical luminosities and inferred accretion rates 5--100 times
lower (e.g. see the compilation of source luminosities by
\pcite{fovame2000}). The three branches of the Z traced out in a CD
or HID by Z sources are called (from top to bottom); Horizontal
Branch, Normal Branch, and Flaring Branch.\newline
\noindent
Studies of the X--ray variability of Z sources revealed two types of
quasi--periodic oscillations (QPOs) with frequencies less than 100 Hz
(Horizontal Branch oscillations; HBO and Normal Branch oscillations;
NBO), twin kHz QPO peaks, and three types of rapid flickering
(``noise''), the very low--frequency noise (VLFN), the low--frequency
noise (LFN), and the high--frequency noise (HFN) (see \pcite{va1995b};
\pcite{va2000} for reviews). The properties of noise features and the
HBOs, such as the central frequency and the fractional rms amplitude,
are strongly correlated with the position of a source along the
Horizontal Branch. These correlations, together with the observed
increase in the ultra--violet flux in the Z source Sco~X--1
(\pcite{1991ApJ...376..278V}) when the source moves from the
Horizontal Branch via the Normal Branch to the Flaring Branch lend
support to the idea that the mass accretion rate increases from the
Horizontal Branch via the Normal Branch to the Flaring Branch (see
\pcite{va1995b}). However, recent observations
(\pcite{1996ApJ...469L...5W}; \pcite{distro2000};
\pcite{homanetal2001}), together with problems noted before (most
notably secular motion of the Z track; \pcite{1994A&A...289..795K})
show that the situation may be more complex (e.g. van der Klis
2001).\newline
\noindent
KHz QPOs have now been seen in just over twenty LMXBs, including all Z
sources (\pcite{va1998}; \pcite{va2000}). Potentially they can provide
a key to measure the basic properties of neutron stars (spin rates and
perhaps magnetic field strengths, radii and masses) and thereby
constrain the equation of state of ultra--dense matter, and to verify
untested general relativistic effects by tracing spacetime just above
the neutron star surface (e.g. \pcite{1993A&AS...97..265K};
\pcite{milaps1998}; \pcite{1997ApJ...480L..27K};
\pcite{1997ApJ...482L.167Z}; \pcite{zhsmst1998}; \pcite{stvi1998};
\pcite{stvi1999}; \pcite{psno2001}). The discovery of kHz QPOs in
GX~5--1 was reported by \scite{1998ApJ...504L..35W}. The HBOs were
discovered by \scite{1985Natur.316..225V}, and the NBOs in GX~5--1 by
\scite{1992MNRAS.256..545L}. The source was also detected at radio
(\pcite{braesmiley72}) and infrared wavelengths
(\pcite{2000MNRAS.315L..57J}).
\noindent
In this paper we present an analysis of all {\it RXTE} observations of
GX~5--1 obtained before 2001. We show for the first time the kHz QPO
separation frequency is not constant in GX~5--1. We show that besides
the two harmonically related low--frequency QPOs found previously on
the Horizontal Branch, two additional harmonically related broad
Lorentzian components are needed to obtain a good fit. We discuss the
power spectra and colour--colour diagrams of the black hole candidate
XTE~J1550--564, the Z source GX~5--1, and the atoll source
4U~1608--52, and conclude that there are some remarkable similarities
and differences which have not been appreciated in the past due to
differences in analysis conventions between in particular neutron
stars and black hole candidates.

\section{Observations and analysis}
\label{analysis}
\noindent
GX~5--1 was observed 76 times in the period spanning July 27 1996 to
March 3 2000 with the proportional counter array (PCA:
\pcite{jaswgi1996}) on board the {\it Rossi X--ray Timing Explorer
(RXTE)} satellite \cite{brrosw1993}. A log of the observations is
presented in Table ~\ref{obs_log}. The total observing time was
$\sim$564 ksec.  During $\sim$40\% of the time only a subset of the 5
detectors was active. Two short observations (2 and 4) were omitted
from our analysis due to data overflows.  Data were always obtained in
a mode providing 16~s time--resolution and a high spectral resolution
(129 channels covering the effective 2--60 keV range; the Standard 2
mode). In addition, a variety of high time--resolution modes was used
in the various observing campaigns; for all observations data with a
time resolution of at least $2^{-11}$~s were obtained in the energy
band spanning 2--60 keV.

\begin{table*}

\caption{Log of the observations of GX~5--1 used in this
analysis.  }
\label{obs_log}

\begin{center}
{\scriptsize
\begin{tabular}{cccc||cccc}
\hline
Obs. & Observation & Date \& &
Amount of good & Obs. & Observation & Date \& &
Amount of good  \\

No. & ID & Start time (UTC) & data (ksec) &
No. & ID & Start time (UTC) & data (ksec)\\
\hline
\hline

1 & 10257-05-01-00  &	21-07-1996	23:42 & $\sim$ 0.14 &39 & 30042-01-18-00 &   21-11-1998      22:35 & $\sim$ 12.0\\
2 & 10257-05-02-00  &	24-10-1996	20:02 & Omitted     &40 & 30042-01-19-00 &   22-11-1998      06:35 & $\sim$ 6.8 \\
3 & 10061-02-01-00  &	02-11-1996	08:34 & $\sim$ 14.5 &41 & 30042-01-20-00 &   22-11-1998      09:47 & $\sim$ 3.5 \\
4 & 10063-01-01-00  &	03-11-1996	20:10 & Omitted     &42 & 40018-02-01-05 &   01-03-2000      14:26 & $\sim$ 1.0 \\
5 & 10063-02-01-00  &	03-11-1996	20:30 & $\sim$ 10   &43 & 40018-02-01-00 &   01-03-2000      23:29 & $\sim$ 26.0\\
6 & 10061-02-02-00  &	06-11-1996	21:19 & $\sim$ 14.8 &44 & 40018-02-01-10 &   02-03-2000      04:29 & $\sim$ 2.3 \\
7 & 10061-02-03-00  &	16-11-1996	00:55 & $\sim$ 15.4 &45 & 40018-02-01-02 &   02-03-2000      06:16 & $\sim$ 5.5 \\
8 & 20055-01-01-00  &	15-02-1997	08:32 & $\sim$ 4.8  &46 & 40018-02-01-03 &   02-03-2000      12:14 & $\sim$ 2.6 \\
9 & 20055-01-02-00  &	12-04-1997	19:00 & $\sim$ 4.8  &47 & 40018-02-01-06 &   02-03-2000      14:05 & $\sim$ 1.9 \\
10 & 20055-01-03-00  &   29-05-1997	19:18 & $\sim$ 5.3  &48 & 40018-02-01-01 &   02-03-2000      15:28 & $\sim$ 2.7 \\
11 & 20053-02-01-00  & 	30-05-1997	09:28 & $\sim$ 17.6 &49 & 40018-02-01-04 &   02-03-2000      17:11 & $\sim$ 14.6\\
12 & 20053-02-01-04 &	06-06-1997	00:28 & $\sim$ 5.2  &50 & 40018-02-02-00 &   03-03-2000      00:00 & $\sim$ 7.6 \\
13 & 20053-02-02-00 &	25-07-1997	05:19 & $\sim$ 9.2  &51 & 40018-02-02-02 &   03-03-2000      07:24 & $\sim$ 13.9\\
14 & 20053-02-01-02 &	25-07-1997	11:43 & $\sim$ 13.1 &52 & 40018-02-02-10 &   03-03-2000      13:48 & $\sim$ 2.7 \\
15 & 20053-02-01-03 &	25-07-1997	18:23 & $\sim$ 5.2  &53 & 40018-02-02-03 &   03-03-2000      23:23 & $\sim$ 26.0\\
16 & 20053-02-01-01 &	25-07-1997	21:41 & $\sim$ 8.3  &54 & 40018-02-02-04 &   04-03-2000      05:56 & $\sim$ 17.0\\
17 & 20055-01-04-00 &	28-07-1997	18:21 & $\sim$ 4.8  &55 & 40018-02-02-05 &   04-03-2000      13:51 & $\sim$ 2.3 \\
18 & 20055-01-05-00 &	21-09-1997	12:29 & $\sim$ 4.6  &56 & 40018-02-02-12 &   04-03-2000      15:37 & $\sim$ 1.6 \\
19 & 30042-01-01-00 &	22-08-1998	10:36 & $\sim$ 11.3 &57 & 40018-02-02-21 &   05-03-2000      00:57 & $\sim$ 23.1\\
20 & 30042-01-02-00 &	14-09-1998	00:25 & $\sim$ 2.8  &58 & 40018-02-02-14 &   05-03-2000      04:14 & $\sim$ 2.7 \\
21 & 30042-01-03-00 &	25-09-1998	03:28 & $\sim$ 9.6  &59 & 40018-02-02-01 &   05-03-2000      06:07 & $\sim$ 2.4 \\
22 & 30042-01-04-00 &	08-10-1998	08:18 & $\sim$ 4.6  &60 & 40018-02-02-20 &   05-03-2000      12:02 & $\sim$ 2.4 \\
23 & 30042-01-02-01 &	09-10-1998	03:30 & $\sim$ 5.7  &61 & 40018-02-02-06 &   05-03-2000      13:43 & $\sim$ 2.3 \\
24 & 30042-01-05-00 &	14-10-1998	00:17 & $\sim$ 10.5 &62 & 40018-02-02-08 &   05-03-2000      15:19 & $\sim$ 2.4 \\
25 & 30042-01-06-00 &	26-10-1998	04:58 & $\sim$ 9.7  &63 & 40018-02-02-13 &   05-03-2000      16:58 & $\sim$ 14.8\\
26 & 30042-01-07-00 &	30-10-1998	05:23 & $\sim$ 5.9  &64 & 40018-02-02-07 &   06-03-2000      00:57 & $\sim$ 1.9 \\
27 & 30042-01-08-01 &	02-11-1998	06:34 & $\sim$ 3.0  &65 & 40018-02-02-17 &   06-03-2000      02:33 & $\sim$ 3.5 \\
28 & 30042-01-08-00 &	03-11-1998	03:18 & $\sim$ 11.9 &66 & 40018-02-02-22 &   06-03-2000      04:09 & $\sim$ 3.4 \\
29 & 30042-01-09-00 &	04-11-1998	05:01 & $\sim$ 6.1  &67 & 40018-02-02-15 &   06-03-2000      05:45 & $\sim$ 3.5 \\
30 & 30042-01-10-00 &	08-11-1998	06:36 & $\sim$ 2.7  &68 & 40018-02-02-11 &   06-03-2000      07:17 & $\sim$ 3.7 \\
31 & 30042-01-11-01 &	08-11-1998	10:01 & $\sim$ 1.3  &69 & 40018-02-02-23 &   06-03-2000      09:07 & $\sim$ 2.7 \\
32 & 30042-01-11-00 &	09-11-1998	06:38 & $\sim$ 8.6  &70 & 40018-02-01-07 &   06-03-2000      10:30 & $\sim$ 1.9 \\
33 & 30042-01-12-00 &	10-11-1998	00:07 & $\sim$ 10.3 &71 & 40018-02-02-16 &   06-03-2000      13:40 & $\sim$ 2.2 \\
34 & 30042-01-13-00 &	10-11-1998	06:40 & $\sim$ 11.2 &72 & 40018-02-01-08 &   06-03-2000      15:16 & $\sim$ 2.2 \\
35 & 30042-01-14-00 &	11-11-1998	03:16 & $\sim$ 10.4 &73 & 40018-02-02-09 &   06-03-2000      16:52 & $\sim$ 14.9\\
36 & 30042-01-15-00 &	20-11-1998	16:10 & $\sim$ 5.4  &74 & 40018-02-02-18 &   07-03-2000      00:53 & $\sim$ 7.2 \\
37 & 30042-01-16-00 &	21-11-1998	00:10 & $\sim$ 23.4 &75 & 40018-02-02-24 &   07-03-2000      04:04 & $\sim$ 3.6 \\
38 & 30042-01-17-00 &	21-11-1998	11:47 & $\sim$ 13.9 &76 & 40018-02-02-19 &   07-03-2000      05:38 & $\sim$ 11.1\\
\end{tabular}
}
\end{center}
\end{table*}

\begin{figure*}
\leavevmode\epsfig{file=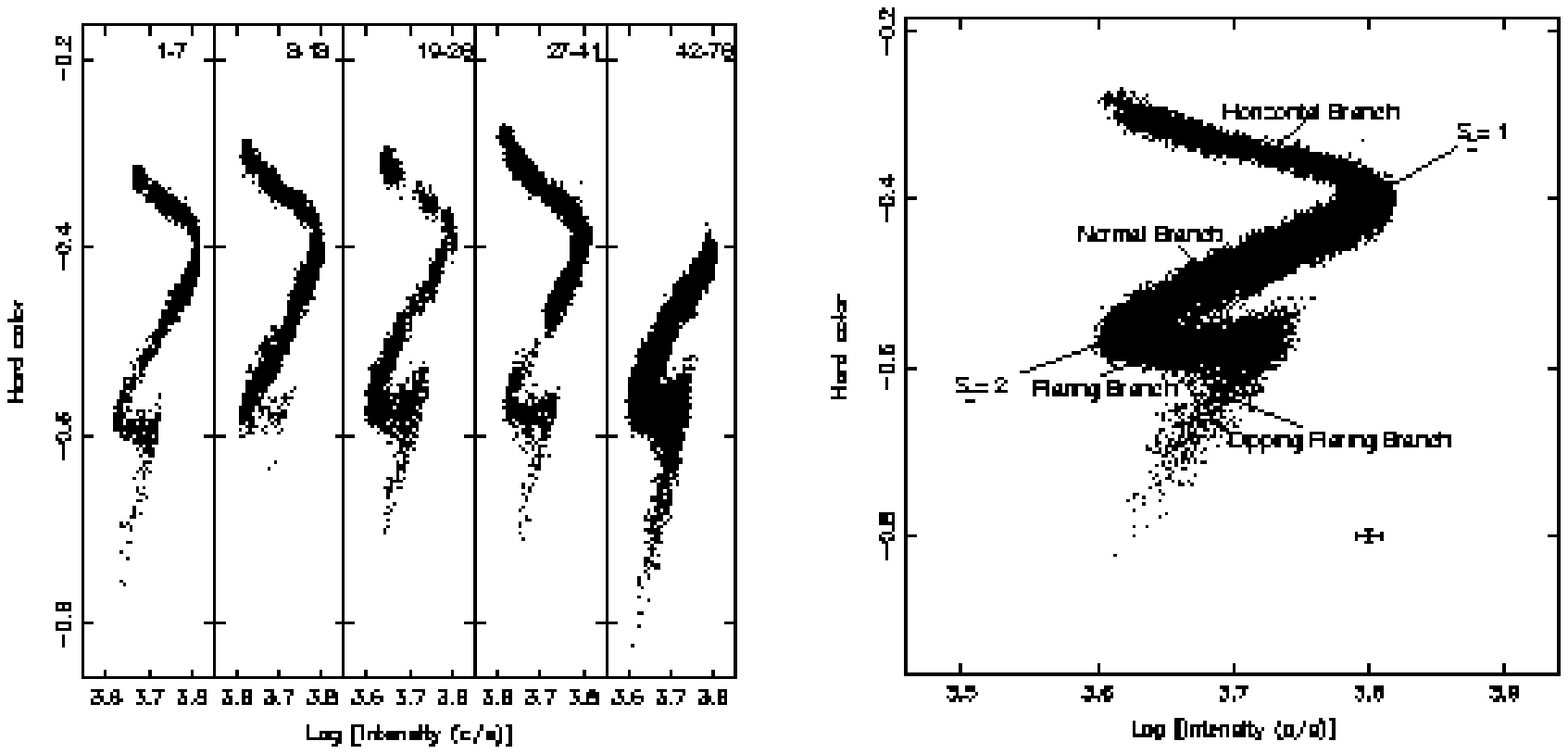,width=9cm,angle=90}\caption{{\it
Left panel:} Hardness--intensity diagrams for observations 1--7,
8--18, 19--26, 27--41, 42--76. {\it Right panel:} Hardness--intensity
diagram of all observations combined. The hard colour is the logarithm
of the 10.2--17.5/6.5--10.2 keV count rate ratio for observations
1--41 and that of the 10.1--17.6 / 6.6--10.1 keV count rate ratio for
observations 42--76. The intensity is the 2.5-17.5 keV (observations
1--41), and the 2.6--17.6 keV (observations 42--76) count rate. The
data were background--subtracted and corrected for changes in the
energy response of the PCA detectors (see text), but no deadtime
corrections were applied (the deadtime correction is $<$4\%). Typical
error bars (not including systematic errors) are shown at the bottom
right of the figure in the right panel.} 
\label{hid}
\end{figure*}

\noindent
From the Standard 2 data we computed HIDs. We only used data obtained
with the two detectors (proportional counter units 0 and 2) that were
always operational since each detector has a slightly different energy
response. The background was subtracted using the standard PCA
background subtraction model version 2.1e, where we used the version
of the {\it skyvle} model appropriate for data from each epoch. The
hardness, or hard colour, and intensity are defined as the logarithm
of the 10.2--17.5 / 6.5--10.2 keV count rate ratio and as the count
rate in the 2.5--17.5 keV band of a 16~s average, respectively (RXTE
energy channels 25--44 / 14--24 and 3--44, respectively). The high
voltage setting of the PCA detectors was changed on March 22 1999
(i.e., the gain changed). Combined with the finite energy resolution
this forced us to use somewhat different energy boundaries in our
computation of the HID for data obtained after March 22 1999 (i.e.,
10.1--17.6 keV / 6.6--10.1 keV for the hard colour and 2.6--17.6 keV
for the intensity for observations 42--76. We used RXTE energy
channels 20--36 / 11--19 and 2--36, respectively). Besides the gain
change the energy response of the detectors changed slowly with
time. These various changes have been partially corrected for using
the Crab pulsar as a reference, under the assumption that the spectrum
of the Crab pulsar does not vary. This correction is only perfect if
the spectral shape of GX~5--1 is equal to that of the Crab (the steps
involved in this correction were outlined in \pcite{dimeva2001} and
\pcite{jonker0918}). So, a small systematic error is introduced both
by the correction for the drift in detector response and the slightly
different energy boundaries which we used as a result of the gain
change. Together with the secular variation of the source
\cite{1996AA...311..197K} this led us to compute 5 separate HIDs
combining observations 1--7, 8--18, 19--26, 27--41, 42--76 (see
Fig.~\ref{hid} {\it left panel}; Table~\ref{obs_log}).

\noindent
In Fig.~\ref{hid} ({\it right panel}) the 5 HIDs are overplotted. The
source was found on the Horizontal Branch, Normal Branch, and Flaring
Branch in four of the five HIDs; for the HID of observations 42--76
the source was not found on the Horizontal Branch. The position of the
source along the Z track is characterized by a parameter called S$_z$
representing curve length along the track (\pcite{havaeb1990};
\pcite{hevawo1992}). The Z tracks were parametrized by fitting a
spline through manually selected points along the track in each of the
5 HIDs separately. The S$_z$ values are determined using the logarithm
of the colours and count rate (\pcite{1997A&A...323..399W}) and
normalized by assigning the hard vertex (the Horizontal Branch--Normal
Branch junction) the value `1' and setting the soft vertex (the Normal
Branch--Flaring Branch junction) to `2'. This defines the measure of
curve--length, and therefore, S$_z = 0$ and S$_z = 3$ are not special
points or vertices. We applied this parameterization to the HID
obtained from each of the subsets. The track traced out by the source
during observations 42--76 only covers part of the Z. In order to
calculate the $S_z$ values along this track we assumed that the hard
vertex in that data set was at the same position relative to the soft
vertex as in the HID of observations 27--41.

\noindent
Using the high time--resolution data we calculated power--density
spectra (2--60 keV) of data stretches of 16~s, up to a Nyquist
frequency of 2048 Hz, using a Fast Fourier Transformation
algorithm. The power spectra were normalized to fractional rms
amplitude squared per Hz, added, and averaged according to their
position along the Z track (the selection method will be described in
detail below). Due to the relatively low amplitude of the variability
in the high frequency part of the power spectrum, simultaneous fits of
the entire frequency range would not constrain the properties of the
high frequency part. Therefore, we fitted the low (1/16--128 Hz) and
high ($\sim$128--2048 Hz) frequency part of the power spectra
separately. The low--frequency part of the average power spectra was
fit with a function consisting of an exponentially cut--off power law
to describe the noise at low frequencies (LFN) and at most five
Lorentzians to fit the QPOs. The component arising in the power
spectrum due to the Poisson counting noise was subtracted. For values
of S$_z$ $>$1.0 a power law was added to the fit function to fit the
very low--frequency noise (VLFN). The function used to describe the
high--frequency part of the average power spectrum was built up out of
two Lorentzians describing the kHz QPOs. Contrary to when we fit the
low--frequency part, the continuum due to Poisson noise was not
subtracted prior to fitting, therefore we included a constant to
account for this Poisson noise. Sometimes also a power law component
was added to account for the slope of the underlying continuum, caused
by the high frequency tail of one of the low--frequency
components. The high frequency tail did not systematically influence
our fit--parameters.

\begin{figure*}
\leavevmode\psfig{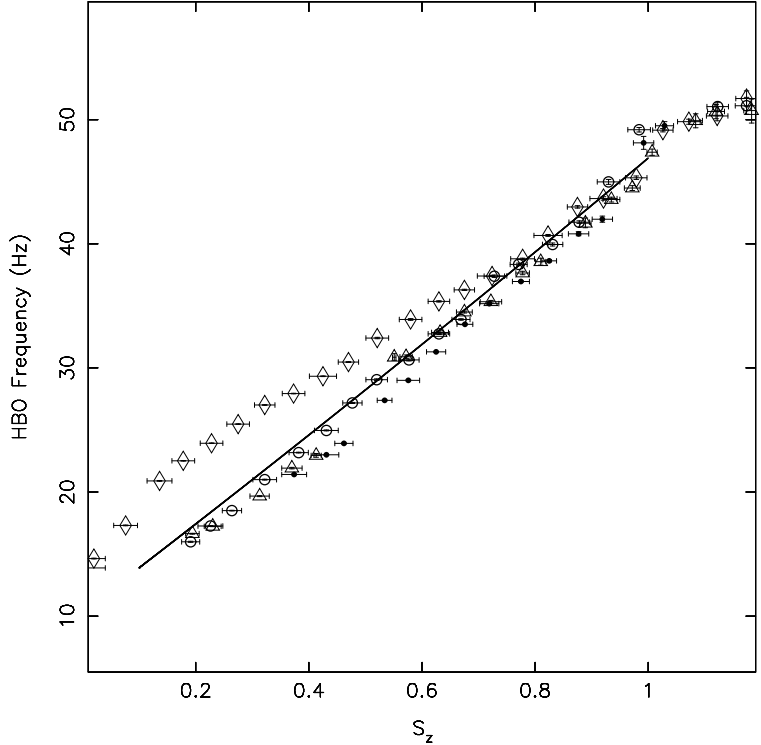}\caption{The
S$_z$--HBO frequency relation measured for four different HIDs. The
S$_z$--HBO frequency relation found for the HID of the observations
27--41 (diamonds) is clearly offset with respect to the relations
found for the other HIDs (dots represent observations 1--7, open
circles observations 8--18, and triangles observations 19--26). The
drawn line is the parent relation to which all the relations are
scaled (see text). Error bars are shown for each measurement; the
error on the HBO frequency is in most cases smaller than the size of
the symbols.}
\label{sz_rel}
\end{figure*}

\noindent
From previous studies of GX~5--1 and other Z sources it is known that
the frequency of the QPOs on the Horizontal Branch (the HBOs and the
kHz QPOs) increases with S$_z$ (\pcite{1998ApJ...504L..35W};
\pcite{2000MNRAS.311..201D}; \pcite{2000ApJ...537..374J};
\pcite{homanetal2001}). For all but the HID of observations 42--76
(when the source was not found on the Horizontal Branch or the upper
part of the Normal Branch and therefore no HBOs were present in the
power spectra; see Fig.~\ref{hid} {\it left panel}) we studied the
relation between S$_z$ and HBO frequency. These relations, plotted in
Fig.~\ref{sz_rel}, have two notable properties. First, the relation
found for observations 27--41 is clearly offset from the other
relations. This is an effect of secular motion of the source during
the gap in between the observations 27--35 and 36--41. During the
first 9 observations (27--35) the source was found on the Horizontal
Branch, whilst, after a 9 day gap, during observations 36--41 the
source traced a part of the Horizontal Branch close to the Normal
Branch, the Normal Branch, and the Flaring Branch. We checked the
vertices of the 5 HIDs to search for additional evidence of secular
motion but in view of the uncertainties involved in their manual
selection, we conclude that the vertices of the 5 HIDs were not
significantly different. The second notable property involves the jump
in HBO frequency near S$_z \sim 1$ (see also plots in
\pcite{1998ApJ...504L..35W}). To investigate the nature of the jump
further we plotted the hard colour vs. the frequency of the HBO. A
discontinuity in the rate of change in frequency as a function of hard
colour was also observed in the hard colour vs. frequency plot at a
frequency of $\sim$50 Hz. Therefore, the jump in frequency when
plotted vs. S$_z$ is not an artefact of the S$_z$ parameterization. We
investigated whether a similar jump was present in the frequencies of
the kHz QPOs, but we lacked the signal--to--noise to conclude on this.

\noindent
In order to combine all power spectra with similar HBO frequencies, we
shifted all S$_z$--HBO frequency relations to one ``parent''
relation. To this end, we fitted a polynomial to the relation between
S$_z$ and HBO frequency for each of the HIDs separately. We included
only measurements for S$_z<1.0$ in the fit. The order of the polynomial
was determined such that the reduced $\chi^2$ of the fit was $\sim$1.
The parent relation for S$_z < 1.0$ is the best linear fit to the
S$_z$--HBO frequency relation for all the HIDs combined ($\nu_{HBO} =
9.9 + 36.8 \times S_{z,parent}$; the drawn line in Fig.~\ref{sz_rel}).
The shifting procedure works as follows: given the S$_z$ value
(S$_{z{\rm ,initial}}$) obtained from the HID corresponding to the
observation the power spectrum is calculated from, the HBO will have a
certain predicted frequency which follows from the S$_z$--HBO
frequency relation found for that HID. That predicted HBO frequency
corresponds, given the S$_z$--HBO frequency parent relation, to a
S$_{z{\rm ,parent}}$. This S$_{z{\rm ,parent}}$ is then assigned to
the power spectrum instead of S$_{z{\rm ,initial}}$ (the change of
S$_z$ corresponds to a horizontal shift in Fig.~\ref{sz_rel}). This
shifting procedure was done for each 16~s power spectrum
separately. Finally, all power spectra were selected according to
their S$_{z{\rm ,parent}}$ values for S$_z<1.0$; the selection bin
width was 0.1. For S$_z > 1.0$ the unshifted values obtained from each
separate HID were used in the selection, since both the secular motion
and the effects of the changes in the response move the Z track in a
direction nearly perpendicular to the Normal Branch. Here the selected
bin width was 0.05. The selected power spectra were averaged and
fitted with the function described above.

\noindent
The errors on the fit parameters were determined using $\Delta\chi^2 =
1.0$ (1$\sigma$, single parameter). The error on S$_z$ is the standard
deviation of the distribution of S$_z$ values in one selection bin.
In cases where components were not significantly detected, 95\%
confidence upper limits were determined using $\Delta\chi^2 =
2.71$. The full--width at half maximum (FWHM) of the Lorentzians was
fixed at 10 Hz in case an upper limit was determined to the sub--HBO,
its third harmonic or the second harmonic of the HBO, and at 30 Hz in
case of the first harmonic (which is the fundamental) of the HBO. The
frequency of the component for which an upper limit was determined was
not fixed but restricted to a range of values around that expected on
the basis of the observed trends. Upper limits to the presence of kHz
QPOs were determined using a FWHM of 75 Hz.

\section{Results}
\label{result}

\subsection{Spectral states}
\begin{figure*}
\leavevmode\psfig{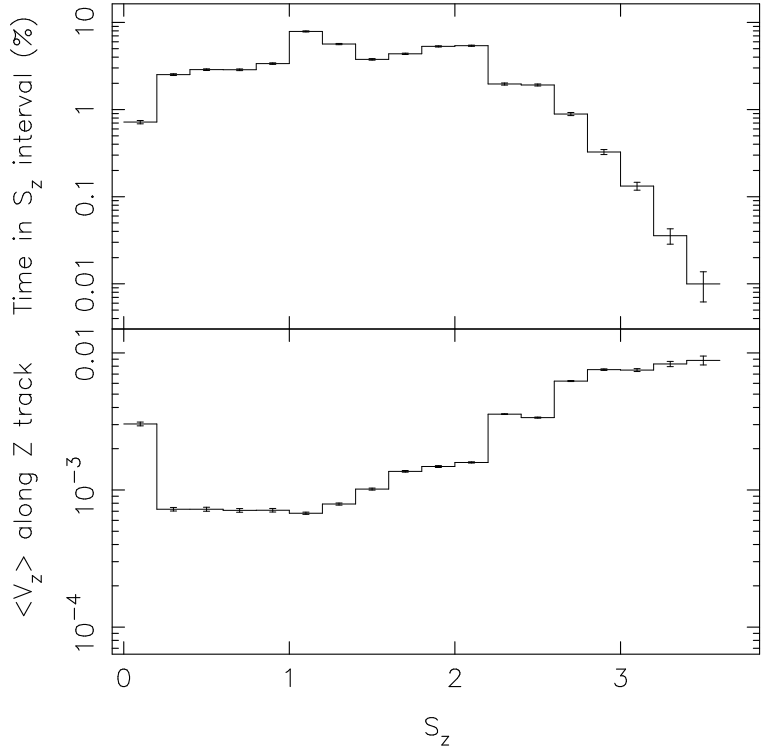}\caption{
{\it Upper panel:} Percentage of the total observing time spent in
each S$_z$ interval as a function of the S$_z$ value. During our
observations the source spent most of the time on the Normal Branch
(S$_z$ values between 1 and 2). {\it Lower panel:}
The average velocity (see text) of the source along the Z track. The
average velocity was approximately constant for S$_z < 1$ (except
towards the lowest S$_z$ values where it increased) but steadily
increased for S$_z > 1 $.}
\label{szinterval} 
\end{figure*}

\noindent
The source was found on the Horizontal Branch, Normal Branch and
Flaring Branch. Note the clear ``Dipping Flaring Branch'' trailing the
Flaring Branch (see Fig.~\ref{hid}, see also
\pcite{1994A&A...289..795K}). GX~5--1 was found to reside most often
on the Normal Branch during the time of our observations (see
Fig.~\ref{szinterval}, {\it upper panel}). The average velocity along
the Z track (defined as the average of ${\rm V_z(i) = \frac
{S_z(i+1)-S_z(i-1)}{T(i+1)-T(i-1)}}$; \pcite{1997A&A...323..399W};
Fig.~\ref{szinterval}, {\it lower panel}) was approximately constant
for $0.1 < $S$_z < 1$ but gradually increased for S$_z > 1 $, i.e., in
anti--correlation with the time the source spent in each part of the Z
diagram, although the product of the two is not exactly constant. Note
that the average velocity along the Z track also increased towards the
lowest S$_z$ values.

\subsection{Low--frequency power spectra}
\begin{figure*}
\leavevmode\psfig{file=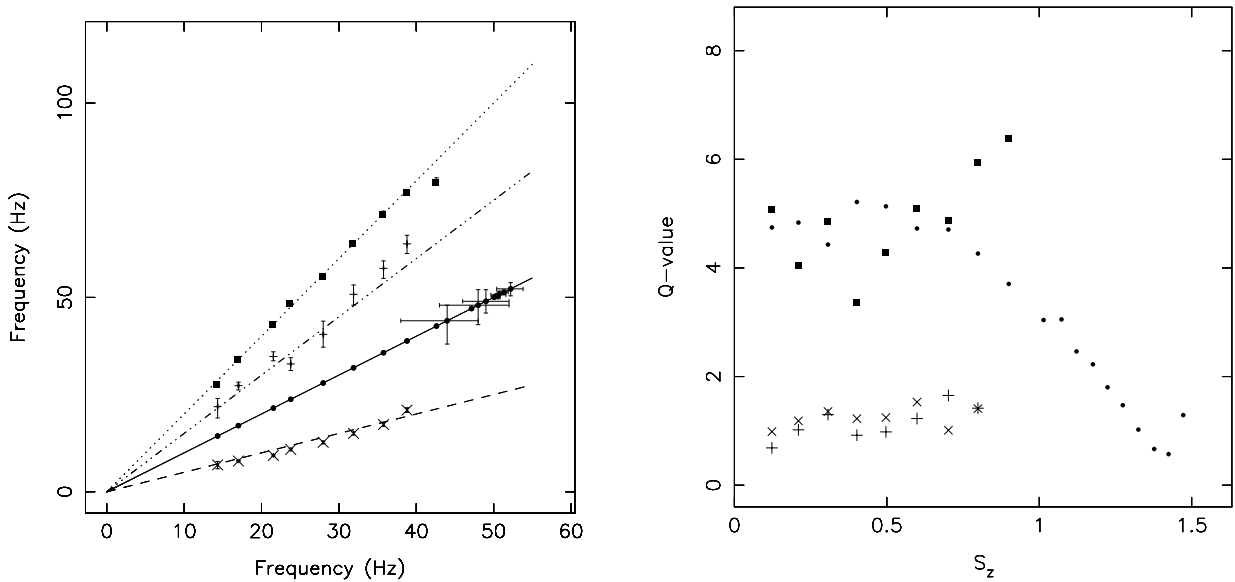,width=8cm,angle=90}\caption{
{\it Left panel:} The frequencies of the four Lorentzian components
used to fit the average power spectra for S$_z < 1.5$, plotted against
the frequency of the HBO. The lines represent 0.5 (dashed line
overlaying the crosses), 1.0 (drawn line overlaying the filled
circles), 1.5 (dashed--dotted line overlaying the plus symbols), and
2.0 (dotted line overlaying the squares) times the HBO
frequency. Error bars are plotted but in several cases they are
smaller than the size of the symbols. {\it Right panel:} The Q--values
of the 4 harmonics. The HBO and its second harmonic have Q--values
near 5 over the range where they are detected simultaneously,
thereafter the HBO coherence drops steadily.  The Q--values of the
sub--HBO and its third harmonic are around 1, with only a slight
increase between S$_z \sim 0.1$ and S$_z \sim 0.8$. Error bars are
omited for clarity but they are a few times larger than the size of
the symbols. The same symbols as in the {\it left panel} have been
used.}
\label{harmonics}
\end{figure*}

\begin{figure*}
\leavevmode\psfig{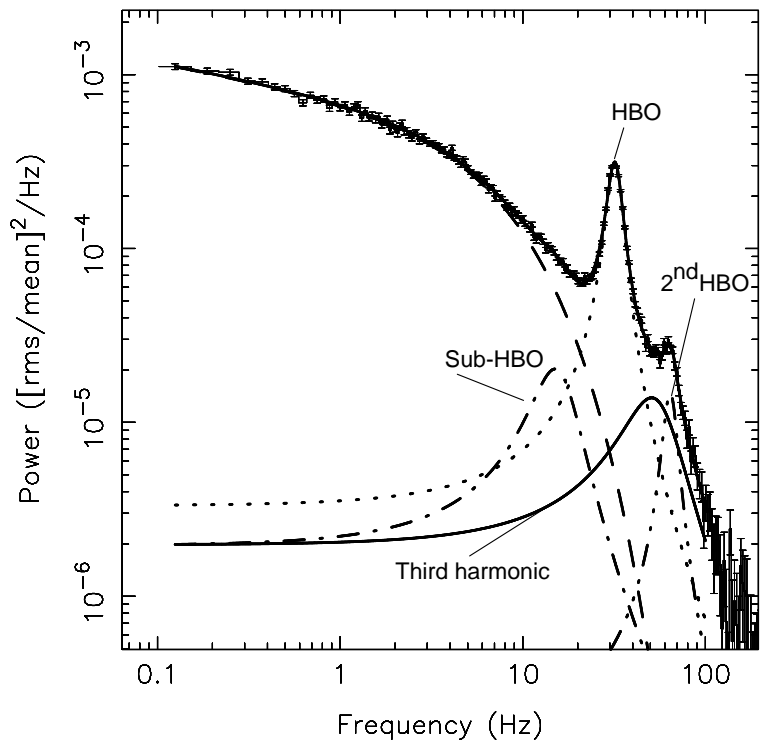}\caption{Average 2--60 keV
power density spectrum for S$_z = 0.60\pm0.03$. The best--fitting and
the individual components used in this fit are indicated; the dashed
line represents the LFN, the dotted line the HBO, the dash--dot line
the sub--HBO, the solid line its third harmonic, and the dash--three
dots line the second hamonic of the HBO. The component arising in the
power spectrum due to Poisson noise was subtracted.}
\label{pds4harm}
\end{figure*}

\noindent
\label{newcomp}
We found two new components in the averaged power spectrum of
GX~5--1. One of these components was located at frequencies consistent
with half the frequency of the HBO, the so--called sub--HBO. This
component was reported before in the Z sources Sco~X--1, GX~340+0, and
GX~17+2 with RXTE (\pcite{wiva1999}; \pcite{2000ApJ...537..374J};
\pcite{homanetal2001}).  The second new component has a frequency
which is consistent with three times the frequency of the sub--HBO (or
1.5 times the frequency of the HBO; see Fig.~\ref{harmonics}). Fitting
the average low--frequency (1/16--256 Hz) power spectrum of S$_z =
0.60\pm0.03$ with a function consisting of a cut--off power law (LFN),
and three Lorentzians (sub--HBO, HBO, and 2${\rm ^{nd}}$ harmonic of
the HBO) a $\chi^2$ of 383 for 209 degrees of freedom was
obtained. Adding a fourth Lorentzian component at frequencies $\sim$3
times the frequency of the sub--HBO gave a $\chi^2$ of 314 for 206
degrees of freedom. An F--test (\pcite{1992drea.book.....B}) to the
$\chi^2$ of the fits with and without the fourth Lorentzian revealed
that the probability that the reduction in $\chi^2$ could be achieved
by a random process is 6.5$\times 10^{-8}$, i.e., the significance of
the addition of the fourth Lorentzian component is $\sim 6
\sigma$. For the other power spectra where this component was found a
similar value was obtained. Although this component is too broad to
qualify formally as a QPO (since the Q--value is less than 2; see
Table~\ref{tabel} and Fig.~\ref{harmonics}, {\it right panel}), for
now we refer to this component as the third harmonic to the sub--HBO
component (see Fig.~\ref{harmonics}). This classification is rendered
strong support by the fact that the Q--values of both the sub--HBO and
its third harmonic are consistent with being the same
(Fig.~\ref{harmonics}; {\it right panel}).
\newline
\noindent
Similar to our work on the Z--source GX~340+0 (Jonker et al. 2000b) we
also experimented with several other fit functions used by various
authors to describe the power spectra of other LMXBs. Both using a fit
function built up out of a broken power law, to fit the LFN component,
and several Lorentzians to fit the QPOs, and a fit function built up
out of just Lorentzians resulted in significantly higher
$\chi^2_{red}$ values than when the fit function described in above
was used ($\chi^2_{red}$= 2.67 for 207 degrees of freedom (d.o.f.),
and $\chi^2_{red}$= 1.93 for 205 d.o.f., respectively), while adding
the fourth Lorentzian component to these fit functions still improved
the $\chi^2_{red}$ significantly (to $\chi^2_{red}$= 2.48 for 204
d.o.f., and $\chi^2_{red}$= 1.73 for 202 d.o.f., respectively).
\newline
\noindent
Besides the two new components, the HBO, its second harmonic and the
LFN were also detected (see Table~\ref{tabel}, \ref{tabel_noise}). A
typical fit (S$_z = 0.60\pm0.03$) showing the contributions of the
individual components is presented in Fig.~\ref{pds4harm}. In four of
the S$_z$ selections the HBO was fit with two Lorentzian peaks in
order to obtain a good fit (see Table~\ref{tabel}). This is either
related to the fact that the HBO moves in frequency within the
selection, or the HBO profile itself is asymmetric. Whenever two
Lorentzian peaks were used to describe the HBO we obtained the rms and
FWHM weighted mean of the two peaks; we used the weighted mean
parameters in Fig.~\ref{harmonics} and in Table~\ref{tabel}. We
weighted the frequencies according to one over the square of the FWHM
and proportionally to the square of the amplitude. The FWHMs were
weighted proportionally to the square of the amplitude. The frequency
offsets between the two peaks were ignored in the weighing since one
of the two Lorentzian peaks contained several times more power than
the other. The powers of the two Lorentzians were added. We would like
to remark that the jump in HBO frequency vs. S$_z$ for each HID
(Fig.~\ref{sz_rel}) has been reduced due to the shifting to
S$_{z,parent}$.

\begin{figure*}
\leavevmode\psfig{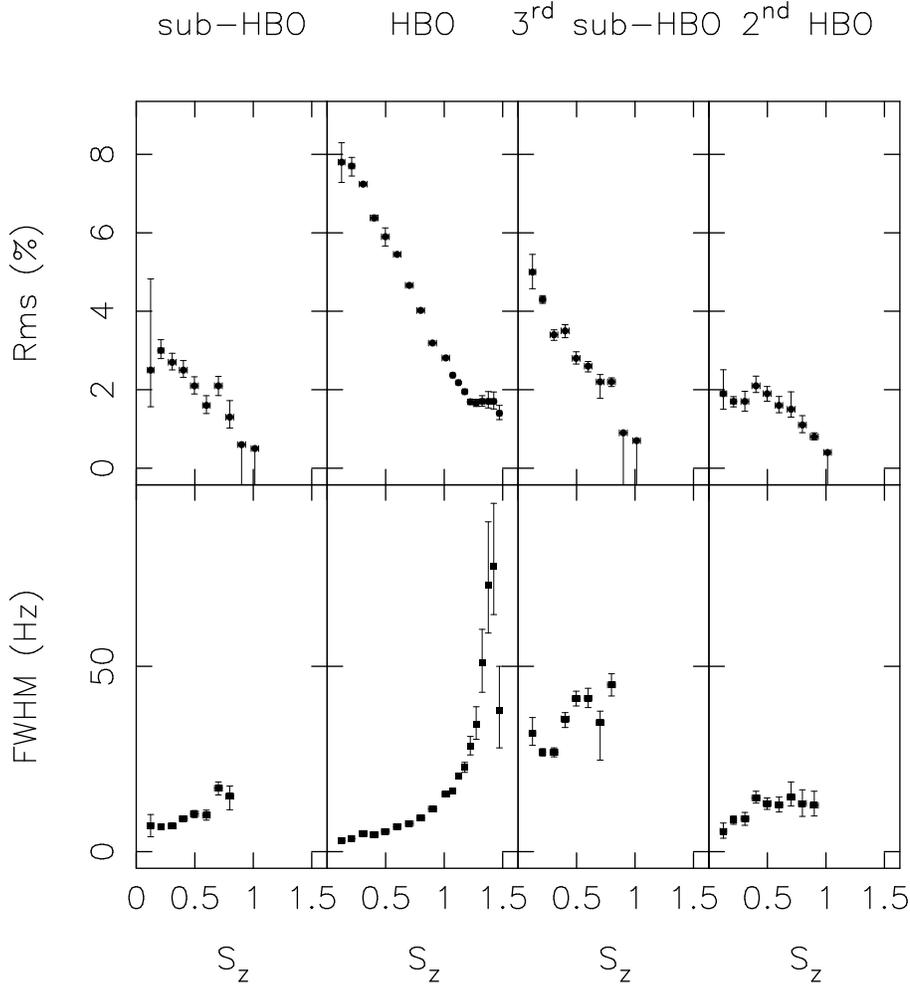}\caption{The
fractional rms amplitude (2--60 keV; top panel) and the FWHM (bottom
panel) of the sub--HBO, the HBO, the third harmonic of the sub--HBO,
and the second harmonic of the HBO as a function of S$_z$. Error bars
are plotted but may be smaller than the size of the symbols. Upper
limits on the fractional rms amplitude of components are plotted using
symbols without a positive error bar. The negative error bar extends
to below zero.}
\label{rms_fwhm}
\end{figure*}

\noindent
The fractional rms amplitude (integrated between frequencies ranging
from 0--$\infty$) of all the low--frequency Lorentzian components
(excluding the NBO) decreased as a function of S$_z$ (see
Fig.~\ref{rms_fwhm} upper panels; Table~\ref{tabel}). For S$_z =
1.0-1.5$ the frequency and the fractional rms amplitude of the HBO was
consistent with being constant at $\sim$50 Hz and $\sim$2\%,
respectively. For S$_z > 1.5 $ the HBO was not detected with an upper
limit of 1.2\%. The FWHM of the Lorentzian components increased as a
function of S$_z$ (see Fig.~\ref{rms_fwhm} lower panels;
Table~\ref{tabel}).

\begin{table*}
\caption{Best--fitting parameters of the Lorentzian components of the
low frequency power spectra (2--60 keV) as a function of S$_z$. }
\label{tabel}
\begin{center}
{\scriptsize
\begin{tabular}{ccccccccccccc}
\hline

 S$_z$ & 
 Sub frac. &  FWHM &  Freq. &
 HBO frac. &  FWHM &  Freq. &
 2$^{nd}_{HBO}$ frac. &  FWHM &  Freq. & 
 3$^{rd}_{sub}$ frac. &  FWHM &  Freq. \\
 value$^c$ &
 rms amp. &  sub &  sub & 
 rms amp. &  HBO &  HBO & 
 rms amp. &  2$^{nd}_{HBO}$ &  2$^{nd}_{HBO}$ & 
 rms amp. &  3$^{rd}$ &  3$^{rd}$\\
  & 
 \% &  (Hz) &  (Hz) &
 \% &  (Hz) &  (Hz) &
 \% &  (Hz) &  (Hz) & 
 \% &  (Hz) &  (Hz)\\

\hline
\hline
0.12$\pm$0.03 & 2.5$^{+1.7}_{-0.8}$ & 7$\pm$3  & 7$\pm$1     & 7.8$\pm$0.5$^b$   & 3.0$\pm$0.2$^b$   & 14.33$\pm$0.05$^b$  & 1.9$\pm$0.6 & 6 $\pm$2 & 27.9$\pm$0.3  & 5.0$\pm$0.3  & 32$\pm$4          & 22$\pm$3 \\
0.21$\pm$0.03 & 3.0$\pm$0.2 &   6.7$\pm$0.4  & 7.9$\pm$0.2   & 7.7$\pm$0.2$^b$   & 3.5$\pm$0.1$^b$   & 17.02$\pm$0.08$^b$  & 1.7$\pm$0.1 & 9 $\pm$1 & 34.4$\pm$0.2  & 4.3$\pm$0.1  & 27$\pm$1	   & 27$\pm$1 \\
0.31$\pm$0.03 & 2.7$\pm$0.2 &   6.9$\pm$0.5  & 9.4$\pm$0.2   & 7.24$\pm$0.03 & 4.9$\pm$0.1   & 21.53$\pm$0.01  & 1.7$\pm$0.3 & 9 $\pm$2 & 43.2$\pm$0.3  & 3.4$\pm$0.1  & 27$\pm$1	   & 35$\pm$1 \\
0.40$\pm$0.03 & 2.5$\pm$0.2 &   8.9$\pm$0.7  & 10.9$\pm$0.2  & 6.38$\pm$0.07$^b$ & 4.6$\pm$0.1$^b$   & 23.78$\pm$0.05$^b$  & 2.1$\pm$0.2 & 15$\pm$2 & 48.8$\pm$0.3  & 3.5$\pm$0.2  & 36$\pm$2	   & 33$\pm$2 \\
0.50$\pm$0.03 & 2.1$\pm$0.2 &   10$\pm$1     & 12.7$\pm$0.3  & 5.9$\pm$0.2 $^b$  & 5.5$\pm$0.2$^b$   & 27.98$\pm$0.06$^b$  & 1.9$\pm$0.2 & 13$\pm$2 & 55.7$\pm$0.2  & 2.8$\pm$0.2  & 41$\pm$2	   & 41$\pm$3 \\
0.60$\pm$0.03 & 1.6$\pm$0.2 &   10$\pm$1     & 15.0$\pm$0.4  & 5.45$\pm$0.04 & 6.8$\pm$0.1   & 31.90$\pm$0.02  & 1.6$\pm$0.2 & 13$\pm$2 & 64.2$\pm$0.3  & 2.6$\pm$0.1  & 41$\pm$3	   & 51$\pm$3 \\
0.70$\pm$0.03 & 2.1$\pm$0.2 &   17$\pm$2     & 17.3$\pm$0.6  & 4.66$\pm$0.03 & 7.6$\pm$0.1   & 35.77$\pm$0.03  & 1.5$\pm$0.3 & 15$\pm$3 & 71.7$\pm$0.5  & 2.2$\pm$0.3  & 35$^{+3}_{-10}$   & 58$\pm$2 \\
0.80$\pm$0.03 & 1.3$\pm$0.3 &   15$\pm$3     & 21.0$\pm$0.6  & 4.02$\pm$0.03 & 9.1$\pm$0.1   & 38.81$\pm$0.03  & 1.1$\pm$0.2 & 13$\pm$4 & 77.2$\pm$0.6  & 2.2$\pm$0.1  & 45$\pm$3	   & 64$\pm$2 \\
0.90$\pm$0.03 & $<$0.6 &      10$^a$ &  	.... &      3.19$\pm$0.02 & 11.5$\pm$0.2  & 42.61$\pm$0.05  & 0.8$\pm$0.1 & 13$\pm$4 & 79.9$\pm$0.9  & $<$0.9       & 10$^a$ &  .... \\
1.01$\pm$0.03 & $<$0.5 &      10$^a$ &  	.... &      2.81$\pm$0.02 & 15.5$\pm$0.3  & 47.14$\pm$0.08  & $<$0.4      &   10$^a$ &  ....      & $<$0.7& 10$^a$ &  ....  \\
1.07$\pm$0.02 & .... &  	.... &  	.... &      2.37$\pm$0.02 & 16.4$\pm$0.4  & 50.0$\pm$0.1    & .... &    .... &  ....      & .... &    .... &  ....   \\
1.12$\pm$0.02 & .... &  	.... &  	.... &      2.18$\pm$0.03 & 20.4$\pm$0.8  & 50.2$\pm$0.2    & .... &    .... &  ....      & .... &    .... &  ....   \\
1.18$\pm$0.02 & .... &  	.... &  	.... &      1.95$\pm$0.04 & 23$\pm$1      & 50.5$\pm$0.3    & .... &    .... &  ....      & .... &    .... &  ....   \\
1.22$\pm$0.02 & .... &  	.... &  	.... &      1.69$\pm$0.07 & 29$\pm$3      & 51.3$\pm$0.5    & .... &    .... &  ....      & .... &    .... &  ....   \\
1.27$\pm$0.02 & .... &  	.... &  	.... &      1.66$\pm$0.09 & 34$\pm$4      & 51$\pm$1        & .... &    .... &  ....      & .... &    .... &  ....   \\
1.33$\pm$0.02 & .... &  	.... &  	.... &      1.7$\pm$0.1   & 51$\pm$9      & 52$\pm$2        & .... &    .... &  ....      & .... &    .... &  ....   \\
1.38$\pm$0.02 & .... &  	.... &  	.... &      1.7$\pm$0.2   & 72$\pm$15     & 48$\pm$5        & .... &    .... &  ....      & .... &    .... &  ....   \\
1.42$\pm$0.02 & .... &  	.... &  	.... &      1.7$\pm$0.2   & 77$\pm$15     & 44$\pm$5        & .... &    .... &  ....      & .... &    .... &  ....   \\
1.47$\pm$0.02 & .... &  	.... &  	.... &      1.4$\pm$0.2  & 38$\pm$11     & 49$\pm$3        & .... &    .... &  ....      & .... &    .... &  ....   \\
\end{tabular}
}
\end{center}
{\footnotesize$^a$Parameter fixed at this value}
{\footnotesize$^b$Parameter is the weighted average of two
Lorentzians (see text)}
{\footnotesize $^c$The error on S$_z$ is the standard deviation of the
distribution of S$_z$ values in the selection bin}
\end{table*}

\begin{table*}

\caption{Best--fitting parameters of the cut--off power law, the power
law, and the NBO component of the low frequency power spectra (2--60
keV) as a function of S$_z$. }
\label{tabel_noise}
\begin{center}
{\scriptsize
\begin{tabular}{ccccccccc}
\hline

S$_z$ value$^c$ &  LFN frac. &  Power law &  Cut--off & VLFN frac. &
Power law &  NBO frac. &  FWHM &  Freq. \\ 
  & rms amp. &  index &  freq. &  rms amp. &  index &  rms amp. &  NBO
&  NBO \\ 
  & \% &   &  (Hz) &  \% &   &  \% &  (Hz) &  (Hz) \\

\hline
\hline
0.12$\pm$0.03 & 7.4$\pm$0.3    & 7e-2$\pm$5e-2  & 2.9$\pm$0.5   & .... &	  .... &	  .... &	  .... &	  .... \\
0.21$\pm$0.03 & 7.57$\pm$0.06  & 2e-2$\pm$1e-2  & 2.97$\pm$0.1  & .... &	  .... &	  .... &	  .... &	  .... \\
0.31$\pm$0.03 & 7.29$\pm$0.06  & 1e-2$\pm$2e-2  & 3.43$\pm$0.1  & .... &	  .... &	  .... &	  .... &	  .... \\
0.40$\pm$0.03 & 7.06$\pm$0.06  & 7e-2$\pm$1e-2  & 4.32$\pm$0.2  & .... &	  .... &	  .... &	  .... &	  .... \\
0.50$\pm$0.03 & 6.83$\pm$0.05  & 0.11$\pm$1e-2   & 5.4$\pm$0.2   & .... &	  .... &	  .... &	  .... &	  .... \\
0.60$\pm$0.03 & 6.48$\pm$0.05  & 0.20$\pm$1e-2   & 7.4$\pm$0.3   & .... &	  .... &	  .... &	  .... &	  .... \\
0.70$\pm$0.03 & 5.42$\pm$0.07  & 0.26$\pm$1e-2   & 8.7$\pm$0.5   & .... &	  .... &	  .... &	  .... &	  .... \\
0.80$\pm$0.03 & 4.81$\pm$0.09  & 0.37$\pm$1e-2   & 15$\pm$1      & .... &	  .... &	  .... &	  .... &	  .... \\
0.90$\pm$0.03 & 4.65$\pm$0.02  & 0.48$\pm$1e-2   & 68$\pm$2      & $<0.4$ &	  1$^a$ &	  $<$0.4 &	  10$^a$ &	  .... \\
1.01$\pm$0.03 & 3.38$\pm$0.02  & 0.41$\pm$1e-2   & 59$\pm$2      & $<0.5$ &	  1$^a$ &	  $<$0.7 &	  10$^a$ &	  .... \\
1.07$\pm$0.02 & $<0.5$ &	0.4$^a$ &	.... &  0.92$\pm$0.07         & 1.0$\pm$0.1  & 1.69$^{+0.05}_{-0.09}$ & 14.4$\pm$0.9    & 6.1$\pm$0.3 \\
1.12$\pm$0.02 & $<0.5$ &	0.4$^a$ &	.... &  1.0$\pm$0.1		 & 1.1$\pm$0.1  & 1.80$\pm$0.04 & 	   9.1$\pm$0.4	   & 6.03$\pm$0.09 \\
1.18$\pm$0.02 & .... &	.... &	.... &  1.0$\pm$0.2		 & 1.3$\pm$0.1  & 2.04$\pm$0.02 &	   7.0$\pm$0.2	   & 5.97$\pm$0.05 \\
1.22$\pm$0.02 & .... &	.... &	.... &  1.3$\pm$0.3		 & 1.4$\pm$0.1  & 2.14$\pm$0.02 &	   5.7$\pm$0.1	   & 5.79$\pm$0.04 \\
1.27$\pm$0.02 & .... &	.... &	.... &  3.1$\pm$0.1		 & 1.9$^a$      & 2.35$\pm$0.02 &	   4.8$\pm$0.1	   & 5.67$\pm$0.03 \\
1.33$\pm$0.02 & .... &	.... &	.... &  1.7$^{+0.6}_{-0.4}$	 & 1.6$\pm$0.2  & 1.9$\pm$0.1  &	   2.7$\pm$0.2	   & 5.25$\pm$0.05 \\
1.38$\pm$0.02 & .... &	.... &	.... &  1.4$\pm$0.3		 & 1.4$\pm$0.1  & 2.0$\pm$0.2  &	   2.3$\pm$0.2	   & 5.33$\pm$0.06 \\
1.42$\pm$0.02 & .... &	.... &	.... &  1.8$\pm$0.4		 & 1.5$\pm$0.1  & 2.23$\pm$0.08 &	   2.6$\pm$0.1	   & 5.43$\pm$0.04 \\
1.47$\pm$0.02 & .... &	.... &	.... &  2.6$\pm$0.1		 & 1.6$^a$      & 2.1$\pm$0.1  &	   2.3$\pm$0.2	   & 5.48$\pm$0.04 \\
1.52$\pm$0.02 & .... &	.... &	.... &  2.2$\pm$0.5		 & 1.5$\pm$0.1  & 2.38$\pm$0.03 &	   3.3$\pm$0.1	   & 5.75$\pm$0.04 \\
1.58$\pm$0.02 & .... &	.... &	.... &  2.2$\pm$0.1		 & 1.5$^a$      & 2.17$\pm$0.03 &	   3.7$\pm$0.2	   & 5.92$\pm$0.05 \\
1.63$\pm$0.03 & .... &	.... &	.... &  2.1$\pm$0.1		 & 1.5$^a$      & 2.00$\pm$0.03 &	   4.0$\pm$0.2	   & 6.01$\pm$0.05 \\
1.68$\pm$0.02 & .... &	.... &	.... &  1.9$\pm$0.3		 & 1.3$\pm$0.1  & 1.84$\pm$0.04 &	   4.2$\pm$0.2	   & 6.07$\pm$0.06 \\
1.73$\pm$0.03 & .... &	.... &	.... &  2.5$\pm$0.4		 & 1.5$\pm$0.1  & 1.61$\pm$0.04 &	   4.3$\pm$0.3	   & 6.09$\pm$0.08 \\
1.82$\pm$0.03 & .... &	.... &	.... &  3.0$\pm$0.2		 & 1.5$\pm$0.1  & 1.30$\pm$0.03 &	   6.2$\pm$0.5	   & 6.2$\pm$0.1 \\
1.95$\pm$0.05 & .... &	.... &	.... &  2.8$\pm$0.2		 & 1.5$\pm$0.1  & 1.14$\pm$0.05 &	   8.1$\pm$0.8	   & 5.8$\pm$0.2 \\
2.05$\pm$0.05 & .... &	.... &	.... &  1.6$\pm$0.1		 & 1.1$\pm$0.1  & 1.68$\pm$0.07$^b$ &	   -0.9$\pm$0.2$^b$ & 4.3$\pm$0.6$^b$ \\
2.15$\pm$0.05 & .... &	.... &	.... &  4.2$\pm$0.2		 & 1.8$^a$      & 2.07$\pm$0.05$^b$ &	   0.3$\pm$0.1$^b$  & 18$\pm$3$^b$ \\
2.36$\pm$0.08 & .... &	.... &	.... &  11$\pm$2		 & 2.0$\pm$0.1  & 1.64$\pm$0.05$^b$ &	   0.4$\pm$0.1$^b$  & 25$\pm$5$^b$ \\
2.64$\pm$0.06 & .... &	.... &	.... &  17$\pm$2
& 1.9$\pm$0.1  & 1.54$\pm$0.07$^b$ &	   0.2$\pm$0.1$^b$  &
25$^{+8}_{-5}$ $^b$ \\
2.90$\pm$0.08 & .... &	.... &	.... &  18$\pm$2		 & 1.8$\pm$0.1  & $<$1.2 &	  0.3$^a$ &	  .... \\
3.13$\pm$0.15 & .... &	.... &	.... &  22$\pm$3		 & 1.9$\pm$0.1  & .... &	  .... &	  .... \\
\end{tabular}
}
\end{center}
{\footnotesize $^a$ Parameter fixed at this value}
{\footnotesize $^b$ Fitted with a cut--off power law, FWHM stands for
power law index, frequency stands for cut--off frequency}
{\footnotesize $^c$ The error on S$_z$ is the standard deviation of the
distribution of S$_z$ values in the selection bin}
\end{table*}

\begin{figure*}
\leavevmode\psfig{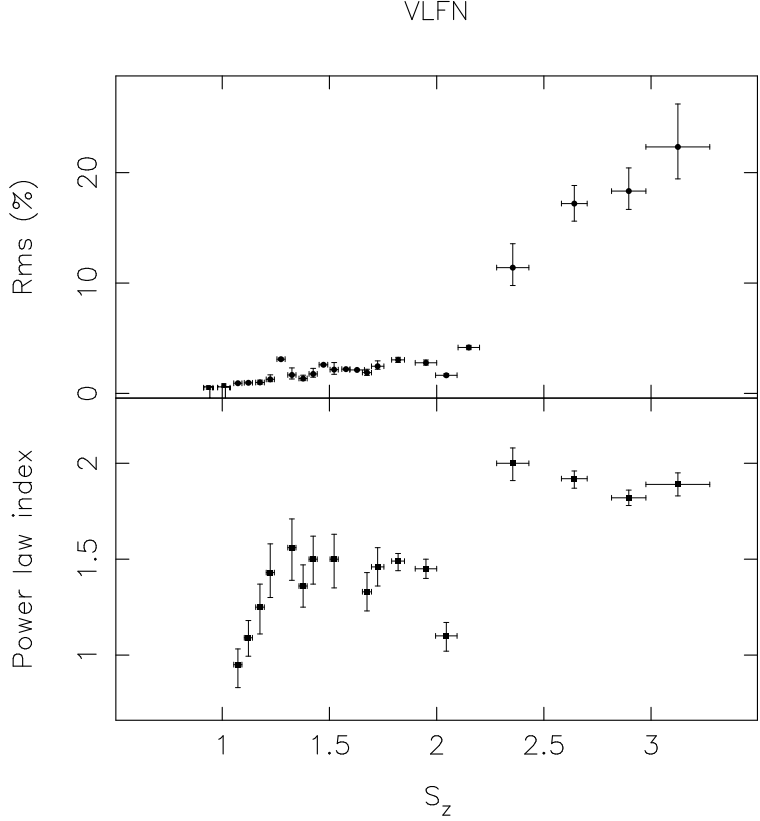}\caption{The
fractional rms amplitude (2--60 keV; top panel) and power law index
(bottom panel) of the VLFN as a function of S$_z$. Error bars are
plotted but may be smaller than the size of the symbols.}
\label{vlfn}
\end{figure*}

\noindent
The fractional rms amplitude of the LFN (0.1--100 Hz) decreased from
7.4\%$\pm$0.3\% to 3.38\%$\pm$0.02\% as S$_z$ increased from
0.12$\pm$0.03 to 1.01$\pm$0.02 (Table~\ref{tabel_noise}). The LFN
component could not be measured for S$_z > 1$. The LFN power law index
increased from 0.07$\pm$0.05 to 0.41$\pm$0.01 while the cut--off
frequency increased from 2.9$\pm$0.5 Hz to $\sim$60 Hz. For S$_z > 1$
we added a power law to the fit function to represent the power at
frequencies below 1 Hz; the very low--frequency noise (VLFN). The VLFN
fractional rms amplitude (0.001--1 Hz) increased from less than 1\% at
the hard vertex to more than 20\% at S$_z > 3.0$. This increase was
gradual for S$_z <2.0$ but steep for $S_z > 2.0 $ (see
Fig.~\ref{vlfn}, Table~\ref{tabel_noise}). The VLFN power law index
gradually increased to 1.5 and remained constant for S$_z < 2.0$, for
S$_z > 2.0$ the power law index was consistent with 2.

\begin{figure*}
\leavevmode\psfig{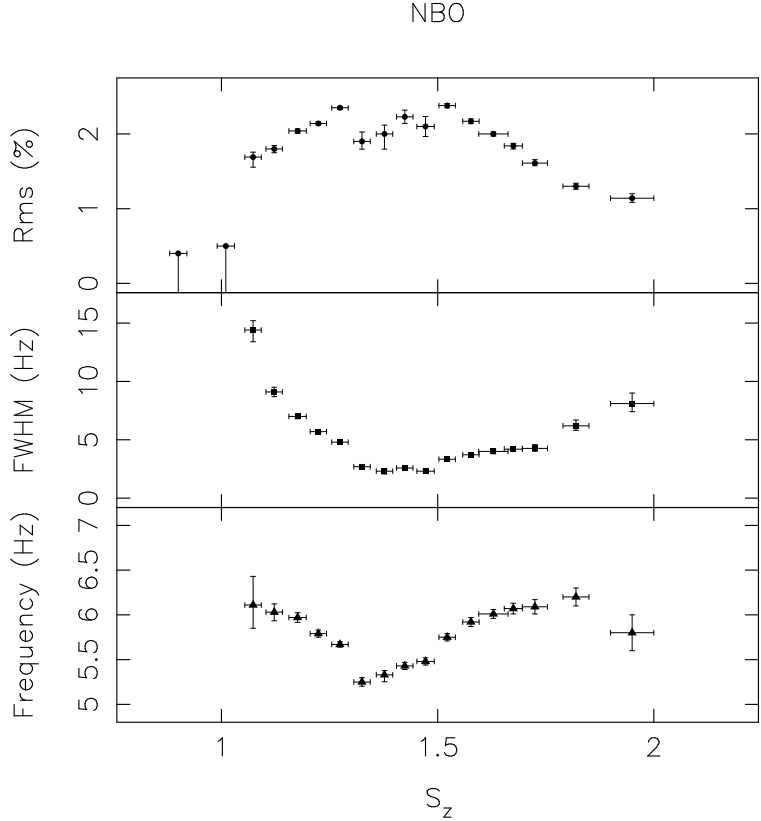}\caption{The
fractional rms amplitude (2--60 keV; top panel), FWHM (middle panel),
and frequency (bottom panel) of the NBO as a function of S$_z$. Error
bars are plotted but may be smaller than the size of the
symbols. Upper limits on the fractional rms amplitude of the NBO are
plotted using symbols without a positive error bar. The negative error
bar extends to below zero.}
\label{nbo_all}
\end{figure*}

\noindent
The frequency of the NBO decreased from 6.1$\pm$0.3 Hz at S$_z =
1.07\pm0.02 $ to 5.25$\pm$0.05 Hz at S$_z = 1.33\pm0.02 $ before it
increased again to 6.1$\pm$0.1 Hz at S$_z = 1.82\pm0.03 $. For S$_z >
1.9$ its frequency could not be determined with high accuracy since
the FWHM increased to 8 Hz. The fractional rms amplitude of the NBO
varied between 1.69\%$^{+0.05}_{-0.09}$\% and 2.38\%$\pm$0.03\% for
S$_z = 1.07\pm0.02$ and S$_z = 1.52\pm0.02$, respectively but it
decreased to $\sim$1.14\%$\pm$0.05\% at S$_z = 1.95\pm0.05$. The
somewhat erratic behaviour in the fractional rms amplitude of the NBO
between S$_z = 1.3-1.5$ can be explained by the fact that the HBO was
very broad and had power in the same frequency range as the
NBO. The fractional rms amplitude is anti--correlated with the
frequency and FWHM of the NBO (see Fig.~\ref{nbo_all}). At S$_z < 1.0
$ stringent upper limits on the presence of the NBO were derived.

\noindent
At S$_z > 2.0$ the NBO was not detected but instead a cut--off power
law was fitted to represent the power at frequencies comparable to the
NBO frequencies. This cut--off power law was peaked (power law index
of $-0.9\pm0.2$) near the soft vertex, with power law index
$-0.9\pm0.2$ and evolved into a flat--topped noise component at S$_z =
2.15\pm0.05$ with a typical power law index of 0.3. The cut--off
frequency increased from 4.3$\pm$0.6 Hz to 25$^{+8}_{-5}$ Hz. The
fractional rms amplitude (0.001-100 Hz) increased from
1.68\%$\pm$0.07\% at S$_z = 2.05\pm0.05$ to 2.07\%$\pm$0.05\% at S$_z
= 2.15\pm0.05$. For S$_z > 2.15$ the fractional rms amplitude was
consistent with 1.6\%. The noise component became undetectable at an
S$_z$ value of 2.64$\pm$0.06 with an upper limit of 1.2\%, where the
vertex between the Flaring Branch and the ``Dipping Flaring Branch''
is at S$_z \sim 2.5$. The fit parameters of the cut--off power law are
marked with a $^b$ in Table~\ref{tabel_noise}.

\subsection{KHz QPOs}
\begin{figure*}
\leavevmode\psfig{file=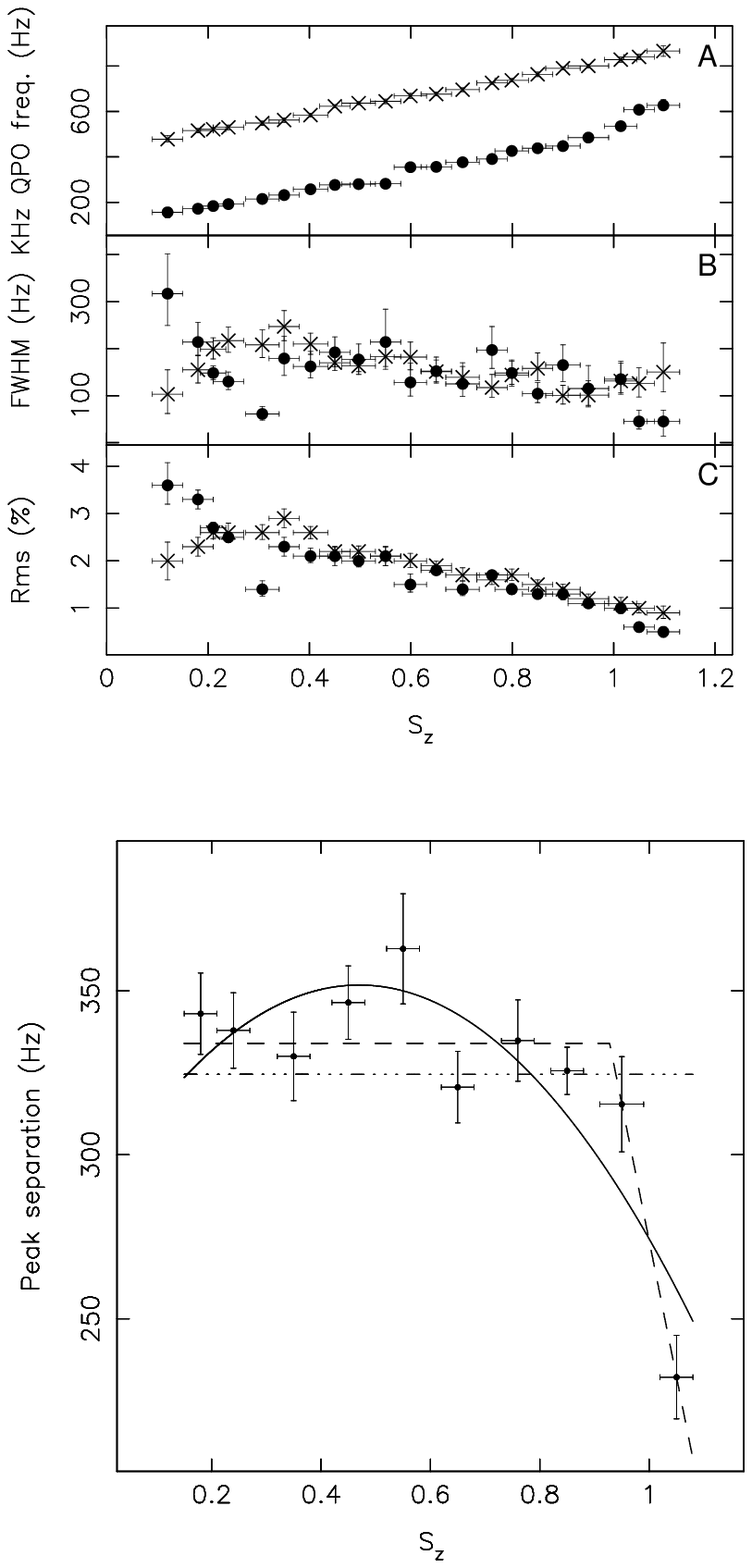,width=8cm}\caption{ The kHz
QPO properties as a function of S$_z$. The measurements presented in
the {\it top panel} are dependent points since data have been used
from partially overlapping bins. ({\it Top panel A:} The frequencies
of the lower (dots) and upper kHz QPO (crosses), the same symbols for
the lower and upper kHz QPO are used throughout the figure. {\it Top
panel B:} The FWHM of the kHz QPO pair. {\it Top panel C:} The
fractional rms amplitude (2--60 keV) of the kHz QPO pair. {\it Bottom
panel:} The peak separation as a function of S$_z$. The dashed--dotted
line represents the best--fitting constant, the drawn line represents the
fit to the data of a parabola, and the dashed line represent a fit
using two straight lines joined at the break with slope fixed to zero
below the break at S$_z \sim 0.93$ (for the fit parameters see
Table~\ref{fits}). Error bars are plotted, but the errors in the {\it
top panel} can be smaller than the size of the symbols.}
\label{khzhbo}
\end{figure*}

\begin{figure*}
\leavevmode\psfig{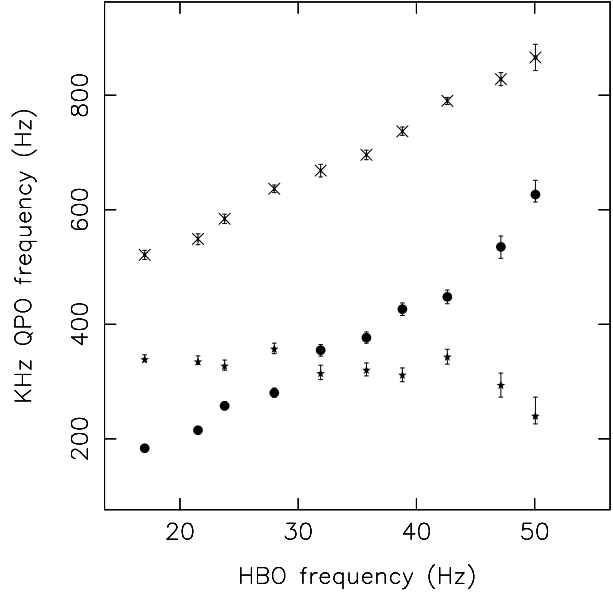}\caption{
The kHz QPO properties as a function of the HBO frequency. The dots
represent the lower kHz QPO and the crosses the upper kHz QPO. The
frequency of the peak separation is indicated with stars. The rate of
increase in lower kHz QPO frequency and therefore also in the peak
separation frequency changes near $\nu_{HBO} \sim 45$ Hz.}
\label{khzhbo2}
\end{figure*}

\noindent
The lower and upper kHz QPO were detected at frequencies ranging from
156$\pm$23 Hz to 627$^{+25}_{-13}$ Hz and from 478$\pm$15 Hz to
866$\pm$23 Hz, respectively (dots and crosses, respectively {\it top
panel}, Fig.~\ref{khzhbo} A). The peak separation ($\Delta \nu$) was
not constant; the fit of a constant to the peak separation (325$\pm$10
Hz) vs. S$_z$ resulted in a $\chi^2_{red}$ of 7 for 10 degrees of
freedom; an unacceptable fit. We tried various other functions
(e.g. see Fig.~\ref{khzhbo} {\it bottom panel} and
Table~\ref{fits}). From an F--test it was clear that the use of a
broken function or a parabola significantly reduced the $\chi^2$
($\sim3.5 \sigma$, $\sim3 \sigma$, respectively) with respect to a
constant, indicating that $\Delta \nu$ is decreasing towards higher
S$_z$ at a rate that is not constant but increases. The data indicate
a decrease in FWHM of the lower kHz QPO towards larger S$_z$, but the
scatter is large. The FWHM of the upper kHz QPO varies between 100 Hz
and 247 Hz (see crosses Fig.~\ref{khzhbo} B {\it top panel}). The
fractional rms amplitude of the lower kHz QPO decreased gradually from
3.6\%$\pm$0.4\% to 0.5\%$\pm$0.1\%, that of the upper kHz QPO first
increased slightly from 2.0\%$\pm$0.4\% to 2.6\%$\pm$0.1\% to decrease
gradually to 0.9\%$\pm$ 0.1\% (see dots and crosses respectively,
Fig.~\ref{khzhbo} C {\it top panel}). The S$_z$ changed from
$\sim0.1-1.1$ over the plotted interval. All fit results are given in
Table ~\ref{khz_prop}.
\newline
\noindent
Motivated by the results on GX~17+2 (\pcite{homanetal2001}) we plotted
the frequency of the kHz QPOs and their separation frequency as a
function of $\nu_{HBO}$ (Fig.~\ref{khzhbo2}). Near $\nu_{HBO} \sim 45$
Hz the peak separation starts to decrease most likely due to an
increase in the rate at which the frequency of the lower kHz QPO
increases with $\nu_{HBO}$. This $\nu_{HBO}$ is consistent with the
jump at $\nu_{HBO} \sim44$ -- $\sim47$ Hz in a $\nu_{HBO}$ vs. S$_z$
plot (see Section~\ref{analysis}).

\begin{table*}
\caption{Best--fitting parameters (2--60 keV) of the lower
($\nu_{low}$) and upper ($\nu_{up}$) kHz QPO as a function of the
source position along the Z track. Note that overlapping selections in
S$_z$ have been used. }
\label{khz_prop}

\begin{center}
{\footnotesize
\begin{tabular}{ccccccc}
\hline
S$_z$ &  Fractional &  FWHM
lower &  Frequency lower &  Fractional &
 FWHM upper &  Frequency upper \\  value$^a$ &
 rms ampli. $\nu_{low}$ &  kHz QPO (Hz) &
 kHz QPO (Hz) &  rms ampli. $\nu_{up}$ &
 kHz QPO (Hz) &  kHz QPO (Hz) \\

\hline
\hline
0.12$\pm$0.03 & 3.6$\pm$0.4 & 317$\pm$77 & 156$\pm$23 & 2.0$\pm$0.4 & 103$\pm$47  & 478$\pm$15\\
0.18$\pm$0.03 & 3.3$\pm$0.2 & 214$\pm$36 & 172$\pm$10 & 2.3$\pm$0.2 & 155$\pm$29  & 515$\pm$9\\
0.21$\pm$0.03 & 2.7$\pm$0.1 & 148$\pm$15 & 183$\pm$3  & 2.6$\pm$0.1 & 199$\pm$23  & 521$\pm$8\\
0.24$\pm$0.03 & 2.5$\pm$0.1 & 130$\pm$19 & 192$\pm$4  & 2.6$\pm$0.2 & 217$\pm$27  & 530$\pm$10\\
0.31$\pm$0.03 & 1.4$\pm$0.2 & 61 $\pm$15 & 215$\pm$4  & 2.6$\pm$0.2 & 208$\pm$30  & 549$\pm$10\\
0.35$\pm$0.03 & 2.3$\pm$0.2 & 179$\pm$42 & 232$\pm$7  & 2.9$\pm$0.2 & 247$\pm$32  & 562$\pm$11\\
0.40$\pm$0.03 & 2.1$\pm$0.2 & 162$\pm$28 & 258$\pm$7  & 2.6$\pm$0.1 & 210$\pm$23  & 584$\pm$8\\
0.45$\pm$0.03 & 2.1$\pm$0.2 & 192$\pm$32 & 277$\pm$9  & 2.2$\pm$0.1 & 170$\pm$18  & 623$\pm$7\\
0.50$\pm$0.03 & 2.0$\pm$0.1 & 177$\pm$29 & 280$\pm$8  & 2.2$\pm$0.1 & 163$\pm$19  & 637$\pm$7\\
0.55$\pm$0.03 & 2.1$\pm$0.2 & 241$\pm$62 & 282$\pm$12 & 2.1$\pm$0.2 & 183$\pm$29  & 645$\pm$10\\
0.60$\pm$0.03 & 1.5$\pm$0.2 & 128$\pm$35 & 355$\pm$10 & 2.0$\pm$0.2 & 182$\pm$30  & 668$\pm$11\\
0.65$\pm$0.03 & 1.8$\pm$0.1 & 152$\pm$28 & 356$\pm$8  & 1.9$\pm$0.1 & 151$\pm$23  & 677$\pm$8\\
0.70$\pm$0.03 & 1.4$\pm$0.2 & 125$\pm$32 & 377$\pm$10 & 1.7$\pm$0.1 & 139$\pm$27  & 696$\pm$8\\
0.76$\pm$0.03 & 1.7$\pm$0.1 & 197$\pm$45 & 391$\pm$19 & 1.6$\pm$0.1 & 117$\pm$25  & 726$\pm$6\\
0.80$\pm$0.03 & 1.4$\pm$0.1 & 148$\pm$26 & 426$\pm$11 & 1.7$\pm$0.1 & 144$\pm$27  & 737$\pm$7\\
0.85$\pm$0.03 & 1.3$\pm$0.1 & 104$\pm$22 & 438$\pm$7  & 1.5$\pm$0.1 & 158$\pm$30  & 764$\pm$10\\
0.90$\pm$0.03 & 1.3$\pm$0.1 & 165$\pm$39 & 448$\pm$12 & 1.4$\pm$0.1 & 100$\pm$20  & 790$\pm$6\\
0.95$\pm$0.04 & 1.1$\pm$0.1 & 115$\pm$43 & 485$\pm$12 & 1.2$\pm$0.1 & 101$\pm$28  & 801$\pm$8\\
1.01$\pm$0.03 & 1.0$\pm$0.1 & 135$\pm$34 & 535$\pm$20 & 1.1$\pm$0.1 & 131$\pm$33  & 828$\pm$11\\
1.05$\pm$0.03 & 0.6$\pm$0.1 & 45 $\pm$20 & 608$\pm$6  & 1.0$\pm$0.1 & 126$\pm$31  & 840$\pm$12\\
1.10$\pm$0.03 & 0.5$\pm$0.1 & 45 $\pm$27 & 627$^{+25}_{-13}$ & 0.9$\pm$0.1 & 150$^{+62}_{-42}$ & 866$\pm$23 \\
1.18$\pm$0.02 & $<$0.7         & 75$^b$         & ....    &      $<$0.9    &  75$^b$         & .... \\
1.22$\pm$0.02 & $<$0.8         & 75$^b$         & ....    &      $<$0.8    &  75$^b$         & .... \\
\end{tabular}
}
\end{center}
{\footnotesize$^a$ The error on S$_z$ is the standard deviation of the
distribution of S$_z$ values in the selection bin}
{\footnotesize$^b$ Parameter fixed at this value}
\end{table*}

\begin{table*}
\caption{Fit parameters of different functions describing the
peak separation as a function of S$_z$. The broken functions are two
straight lines concatenated at the break. The last row contains the
$\chi^2$ and the degrees of freedom (d.o.f.) of the fit. Note that
unlike in the {\it top panel} of Fig~\ref{khzhbo} only independent
measurements have been used in the {\it lower panel}. }
\label{fits}
\begin{center}
{\footnotesize
\begin{tabular}{cccccc}
\hline
  &  Constant & Linear &  Parabola &  Broken &  Broken \\

\hline
\hline
Zeropoint       & 325$\pm$10 Hz & 369$\pm$21 Hz& $291\pm29$ Hz  & 334$\pm$4 Hz    & 348$\pm$9 Hz   \\
Slope            & ....    & $-70\pm30$ & $259\pm109$ & 0$^a$         & -24$\pm$15    \\
Quadratic / break& ....    & ....    & $-275\pm90$ & S$_z = 0.93\pm$0.06 & S$_z = 0.94\pm$0.05 \\
Slope after Break & ....    & ....    & ....     & -832$\pm$483  & -832$\pm$439  \\
$\chi^2 / $d.o.f.& 66.8 /9    & 39.9/8     & 15.0 / 7    & 7.8 / 7       & 5.5 / 6      \\

\end{tabular}
}
\end{center}
{\footnotesize $^a$ Parameter fixed at this value}

\end{table*}

\section{Discussion}

\noindent
A detailed analysis of all {\it RXTE} observations obtained to date of
GX~5--1 was presented. The main result of our high--frequency
variability study is that the kHz QPO peak separation is not constant
in GX~5--1. The low--frequency power spectra for S$_z < 1.0 $ are
complex and can be best described by 4 harmonically related
Lorentzians, in addition to the low--frequency noise (LFN). Two of these
Lorentzians were not known previously in GX~5--1. At S$_z > 1.9 $ a
new component was found in the power spectra. Below we discuss these
findings and compare the timing properties and colour--colour diagrams
of GX~5--1, the black hole candidate XTE~J1550--564, and the atoll
source 4U~1608--52.

\subsection{KHz QPOs}
\noindent
We showed that the kHz QPO peak separation is not constant. The data
is consistent with a constant peak separation followed by a steep
decrease for S$_z \sim 1.0$, or a parabolic relation between the peak
separation and S$_z$. A decrease in peak separation was previously
found in Sco~X--1 (\pcite{vawiho1997}), 4U~1608--52
(\pcite{mevawi1998}), 4U~1728--34 (\pcite{1999ApJ...517L..51M}),
4U~1735--44 (\pcite{1998ApJ...508L.155F}), 4U~1702--43
(\pcite{mastsw1999}), and GX~17+2 (\pcite{homanetal2001}). We
found that in GX~5--1 the rate of increase in frequency of the lower
kHz QPO changes near $\nu_{HBO} \sim 45$Hz, causing the peak
separation to decrease.
\newline
\noindent
The frequencies of the kHz QPOs of GX~5--1 are low in comparison with
the other kHz QPO Z sources (the lowest frequencies of the lower and
upper kHz QPO we found are $156\pm23$ Hz, and $478\pm15$ Hz,
respectively).  The lowest upper kHz QPO frequency found so far in an
atoll source (449$\pm$20 Hz for 4U~0614+09; \pcite{vafova2000}) is
at the low--frequency limit ($\sim$500 Hz for the upper kHz QPO;
\pcite{milaps1998}) imposed by the sonic point beat frequency
model. The cause of the lower bound on the Keplerian frequency in the
sonic point beat frequency model is that the radiation drag of the
luminosity produced near the surface of the neutron star can only
remove the required amount of the specific angular momentum of the gas
to make it fall to the neutron star when the gas is close to the
neutron star (\pcite{milaps1998}). Unlike 4U~0614+09, in the case of
GX~5--1 QPOs with still lower frequency did not occur because the
source was not found at lower S$_z$ values but not because the QPO
disappeared. Therefore, observations of kHz QPOs in GX~5--1 at even
lower S$_z$ values will provide a strong test of the sonic point beat
frequency model.

\begin{figure*}
\leavevmode\psfig{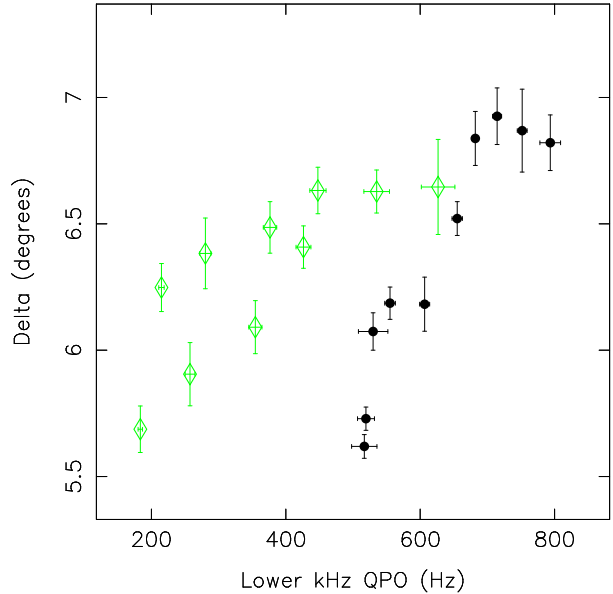}\caption{ The angle
$\delta$ (defined by Osherovich \& Titarchuk 1999) for GX~5--1 (grey
diamonds) and GX~17+2 (black dots) as a function of the lower kHz QPO
frequency.}
\label{delta}
\end{figure*}

\noindent
In the so--called two--oscillator model (\pcite{1998ApJ...499..315T};
\pcite{1999ApJ...523L..73O}) the angle $\delta$ is defined ($\delta =
arcsin[(\nu_{upper}^2-\nu_{lower}^2)^{-0.5} (\nu_{HBO}
\nu_{upper}/\nu_{lower})]$, where $\nu_{lower}$ is the frequency of
the lower kHz QPO, $\nu_{upper}$ is the frequency of the upper kHz
QPO, and $\nu_{HBO}$ is the frequency of the HBO; $\delta$ is the
angle between the magnetosphere equator and the disc plane in the
model). The current version of the theory predicts $\delta$ to be
constant. However, the measurements of \scite{homanetal2001} for
GX~17+2 and the our measurements for GX~5--1 show that $\delta$
vs. the frequency of the lower kHz QPO changes is not constant (see
Fig.~\ref{delta}). A fit of a constant gives $\delta=6.1\pm0.2$ and a
$\chi^2$ of 377 for 9 d.o.f. for GX~17+2 and $\delta=6.3\pm0.1$ and a
$\chi^2$ of 95.2 for 9 d.o.f. for GX~5--1.

\subsection{Comparison with other Z sources}
\begin{figure*}
\leavevmode\psfig{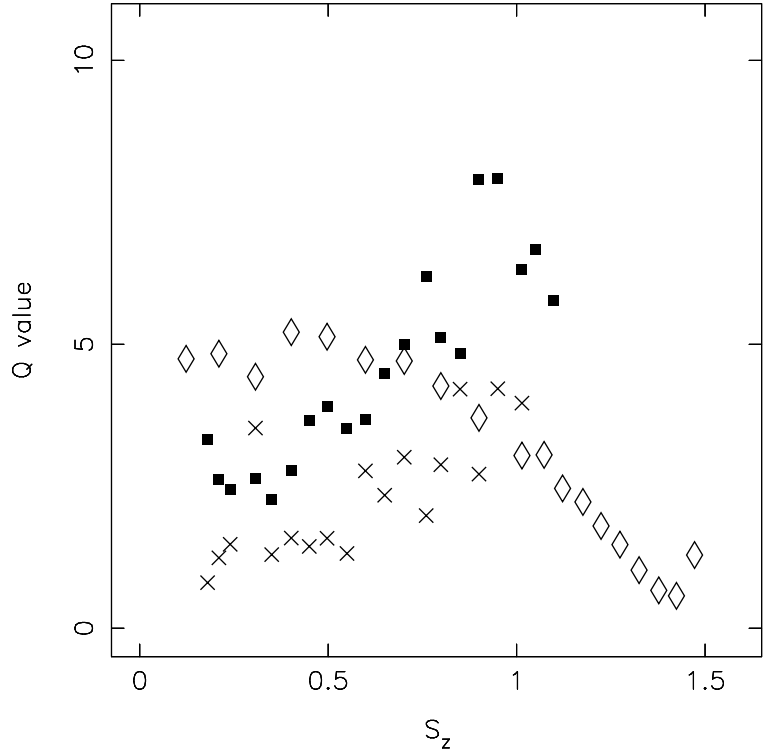}\caption{ The
Q--values of the HBO (diamonds), the lower (crosses), and the upper
kHz QPO (filled squares) as a function of S$_z$. The Q--values of both
kHz QPOs increase with increasing S$_z$, while that of the HBO
decreases for S$_z > 0.5$. Error bars have been omitted for
clarity. The errors of the diamonds (HBO) are $\sim1.5$ times the size
of the symbols. The errors of the crosses and squares (lower and upper
kHz QPO, respectively) are of the order of the amplitude of the
scatter in measurements close together in S$_z$.}
\label{qs}
\end{figure*}

\noindent
We have detected a sub--HBO component similar to the component in the
Z--sources Sco~X--1 (\pcite{wiva1999}), GX~340+0
\cite{2000ApJ...537..374J}, and GX~17+2
(\pcite{homanetal2001}). Furthermore, we found a fourth Lorentzian
component whose frequency was consistent with 1.5 times the frequency
of the HBO. If we interpret the sub--HBO as the fundamental frequency,
then the HBO would be the second harmonic, the newly found peak at 1.5
times the HBO frequency the third harmonic, and the second harmonic of
the HBO the fourth harmonic. In such a scheme the odd harmonics should
either be formed less coherently than the even harmonics, or an
additional broadening mechanism has to be invoked which broadens the
odd harmonics more than the even harmonics (see Fig.~\ref{harmonics};
{\it right panel}). This harmonic structure could be explained in a
scenario where the HBO is caused by a warped accretion disc with a
two--fold symmetry. The warp itself is stable to deviations from its
two--fold symmetry, i.e., variations in the warp take place on a
timescale long compared to the timescale of variations in the
difference between the two sides of the warp, leading to broader odd
harmonics. This connects to the models proposed by \scite{stvi1998}
and \scite{psno2001}. In the magnetic beat--frequency model for the
HBO (\pcite{alsh1985}; \pcite{lashal1985}) the sub--HBO should reflect
differences between the two magnetic poles. Differences in the
magnetic field configuration are not likely to vary in time rapidly,
so the extra broadening of the sub--HBO with respect to the HBO is
unexplained. This necessitates the introduction of an additional
broadening mechanism working on the odd harmonics only after their
formation. 
\newline
\noindent
Prior to the detection of the sub--HBO and its third harmonic, the
frequency of the second harmonic of the HBO was not measured to be
exactly twice the HBO frequency (\pcite{1998ApJ...504L..35W}); with
the introduction of the third harmonic of the sub--HBO in the fit
function this discrepancy has disappeared (see Fig.~\ref{harmonics};
{\it left panel}). Power density spectra calculated from data obtained
with the EXOSAT and Ginga satellites (\pcite{1994A&A...289..795K};
\pcite{1992MNRAS.256..545L}) were fit using only the even harmonics
(the HBO and its 2${\rm^{nd}}$ harmonic) and an additional high
frequency noise (HFN) component. Now, using the larger collecting area
of RXTE we find that the HFN in GX~5--1 is better described by two
Lorentzian components which have centroid frequencies of 0.5 and 1.5
times that of the HBO.
\newline
\noindent
The same function we used for our analysis of GX~5--1 was used in the
analysis of the similar Z source GX~340+0
\cite{2000ApJ...537..374J}. There no third harmonic of the sub--HBO
was found. Instead we found excess power (a shoulder) close to the HBO
peak, which when fit with a Lorentzian was at a frequency entirely
inconsistent with 1.5 times that of the HBO. Reanalysis of those data
with the insights gained by our work on GX~5--1 did not alter this
conclusion. In GX~340+0 both the HBO and the NBO peak became
asymmetric when they were strongest. Similar behaviour was found in
GX~5--1, where for S$_z < 0.5$ four of the five HBO measurements were
done using two Lorentzians with centroid frequencies up to 2 Hz apart
with similar FWHMs but different strengths, indicating that the HBO
was asymmetric. However, unlike in our analysis of GX~340+0 where the
data were selected according to HBO frequency, in the case of GX~5--1
this asymmetry could be an artifact of the selection method as
explained in Section~\ref{newcomp}. An attempt to fit the average
power spectra of GX~17+2 (\pcite{homanetal2001}) with the same fit
function as that of GX~5--1 did not lead to the detection of a third
harmonic to the sub--HBO. The asymmetry of the HBO found in GX~340+0
and the third harmonic of the sub--HBO found in GX~5--1 could be two
different components, one of which is detected in each source because
of the special circumstances affecting detections in each case: the
larger amplitude of the HBO in GX~340+0 than in GX~5--1 allows to
detect a strong asymmetric (shoulder) component, whereas the higher
signal--to--noise in GX~5--1 than in GX~340+0 (due to the fact that
GX~5--1 is several times brighter than GX~340+0) leads to the
detection of the broad third harmonic. New observations with higher
signal--to--noise ratios can reveal whether both components are
present in both sources or whether they represent power provided by
one single component.
\newline
\noindent
The low--frequency power spectra of the three Z sources discussed
above (GX~17+2, GX~340+0, GX~5--1) are quantitatively different.  In
Sco~X--1 and GX~17+2 a QPO is found when the source is on the Flaring
Branch (the FBO; \pcite{1989ApJ...337..843H};
\pcite{1990MNRAS.243..114P}). Such a QPO is absent in GX~340+0,
GX~5--1, GX~349+2, and Cyg~X--2. The new component we found in the
power spectrum of GX~5--1 with frequencies similar to those of the FBO
when the source is on the Flaring Branch is much broader than the FBO
in Sco~X--1 and GX~17+2. The second harmonic of the HBO is relatively
strong in GX~17+2 while for GX~5--1 and GX~340+0 only weak harmonics
have been detected (see \pcite{hava1989};
\pcite{1989ARA&A..27..517V}). Furthermore, the fractional rms
amplitude of the HBO is highest in GX~340+0, and lowest for
GX~17+2. The HBO frequency in GX~17+2 was found to decrease on the
Normal Branch (\pcite{wihova1997}; \pcite{homanetal2001}), whereas in
GX~340+0 and GX~5--1 the frequency is consistent with being constant
(\pcite{2000ApJ...537..374J}). Besides the Lorentzians also the LFN
properties are different; it is always peaked in GX~17+2 whereas this
is never the case in GX~5--1 and GX~340+0 (see \pcite{hava1989};
\pcite{1989ARA&A..27..517V}; \pcite{homanetal2001};
\pcite{2000ApJ...537..374J}). The dependence of the HBO, lower and
upper kHz QPO Q--values on S$_z$ is shown in Fig.~\ref{qs}. The HBO
Q--value decreases gradually for S$_z > 0.5$, while that of the kHz
QPOs increases. This is opposite to what was found for GX~17+2
(\pcite{homanetal2001}), where the Q--value of the HBO and that of the
kHz QPOs increased as a function of S$_z$.

\subsection{Comparing GX~5--1 with XTE~J1550--564 and 4U~1608--52}
\begin{figure*}
\leavevmode\psfig{file=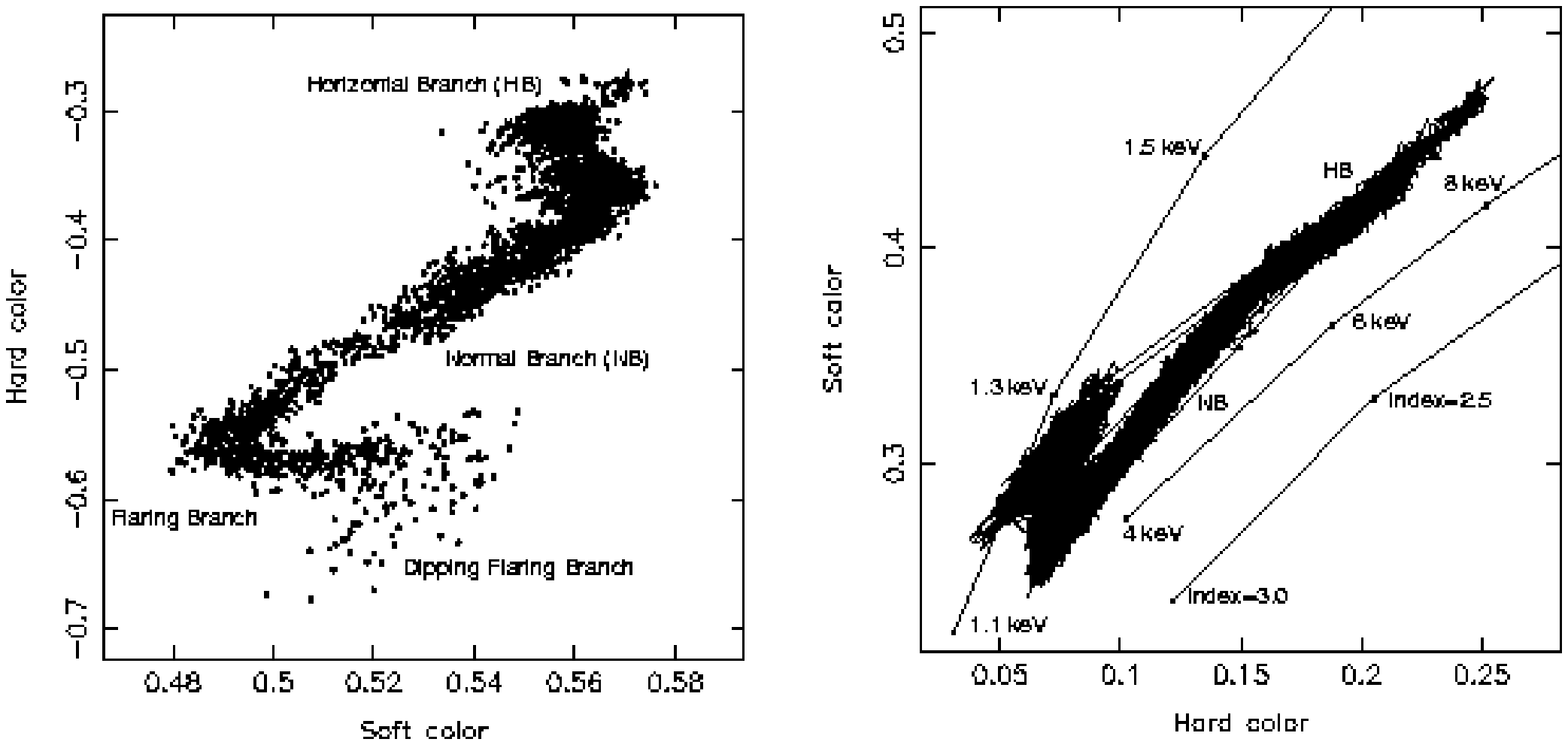,width=9cm,angle=90}
\caption{ {\it Left panel:} Colour--colour diagram of observations
19--41 where the soft and hard colour were defined as the logarithm of
the 3.6--6.2 / 2.5--3.6 keV and 9.8--16.0 / 6.2--9.8 keV count rate
ratio, respectively. Each dot is a 64~s average. {\it Right panel:}
Colour--colour diagram of the 6.4--9.7 / 2--6.4 keV (or 6.6--9.7 /
2--6.6 keV for observations 42--76; soft colour) count rate ratio
vs. the 9.8--16.0 / 2--6.4 keV (or 9.7--15.8 / 2--6.6 keV for
observations 42--76; hard colour). Each 512~s average is connected to
the next by a line. The lines in the {\it right panel} indicate what
RXTE would have seen if the source spectrum had consisted of only a
blackbody (top), a thermal bremsstrahlung (middle), or a powerlaw
(bottom line) spectrum. In this Figure as well as in Figure
~\ref{1550_as_BHC}, and ~\ref{1608_as_BHC} error bars were omitted for
clarity; data were background subtracted but no deadtime corrections
were applied (the deadtime was in all cases less than 4\%).}
\label{z_as_BHC}
\end{figure*}

\begin{figure*}
\leavevmode\psfig{file=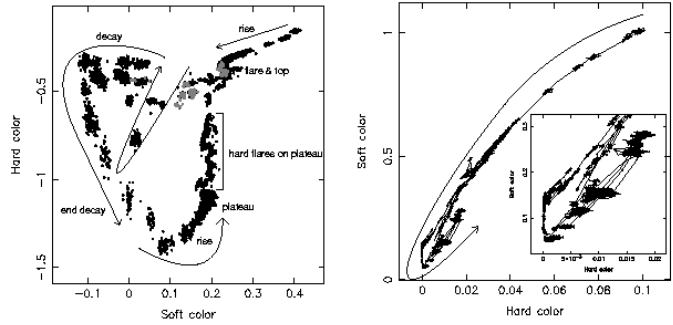,width=9cm,angle=90}
\caption{ {\it Left panel:} Colour--colour diagram of
XTE~J1550--564. We used all data obtained with the RXTE satellite of
XTE~J1550--564 during the interval MJD 51065--51259 (gain epoch 3);
soft and hard colours were defined as in Fig.~\ref{z_as_BHC} ({\it
left panel}). Each dot is a 64~s average. The path the source traces
as the outburst progresses is indicated. The grey dots are the last
days of observations just before the gain change. {\it Right panel:}
Colour--colour diagram of the same data as in the {\it left
panel}. Note that the soft colour is plotted vs. hard colour. The hard
and soft colour were defined as the 16.0--19.4 / 2.2--6.2 keV, and
6.5--15.7 / 2.2--6.2 keV count rate ratio, respectively. Each 64~s
average is connected to the next by a line. A zoom--in of the lower
left corner is visible to the right. The arrow indicates the source
changes as the outburst progressed.}
\label{1550_as_BHC}
\end{figure*}

\begin{figure*}
\leavevmode\epsfig{file=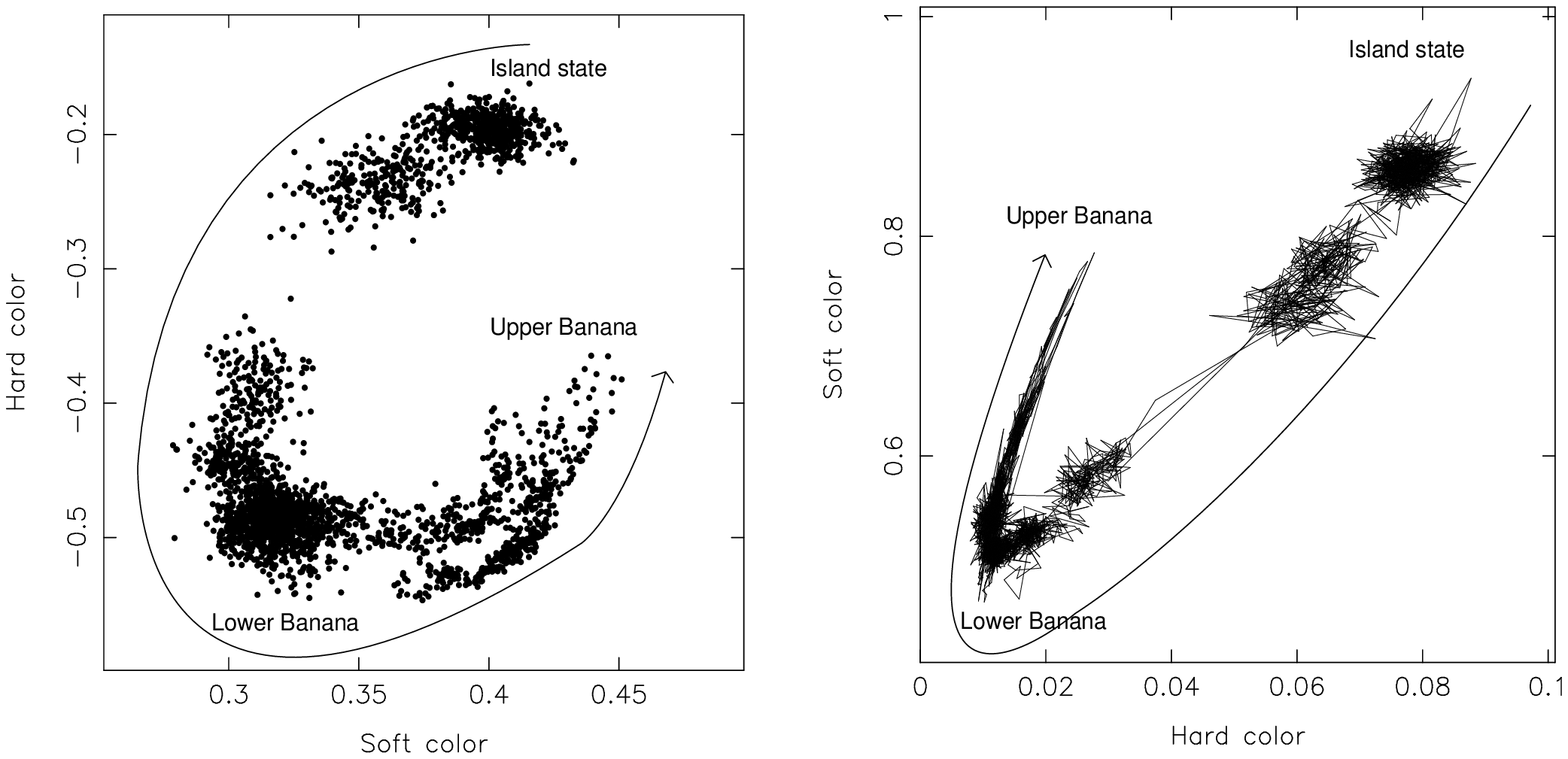,width=9cm,angle=90}
\caption{ {\it Left panel:} Colour--colour diagram of 4U~1608--52. We
used the same data as was used by M\'endez et al. (1998). Soft and
hard colours where defined as in Fig.~\ref{z_as_BHC} ({\it left
panel}).  Each dot is a 64~s average. {\it Right panel:}
Colour--colour diagram of the same data as in the {\it left
panel}. Note that the soft colour is plotted vs. hard colour. The hard
and soft colour were defined as in Fig.~\ref{1550_as_BHC}.  Each 64~s
average is connected to the next by a line. The arrow indicates the
direction of increase in inferred mass accrettion rate.}
\label{1608_as_BHC}
\end{figure*}

\noindent
In the power spectra of black hole candidates low--frequency QPOs with
harmonics are occasionally found in the low, intermediate and very
high state (e.g. GS~1124--68, \pcite{bevale1997}; GX~339--4,
\pcite{meva1997}; XTE~J1550--564, \pcite{somcre2000b};
\pcite{2001ApJS..132..377H}). Occasionally, these black hole
candidate QPOs have asymmetric profiles. We found a similar harmonic
structure of QPOs in the Z source GX~5--1. The Q--values of the
harmonics for XTE~J1550--564 are similar to those found for GX~5--1:
low Q--values for the odd harmonics and high Q--values for the even
harmonics (\pcite{2001ApJS..132..377H}). However, we checked the
energy dependence of the QPOs in GX~5--1 and there are no indications
of changes in the Q--values with energy as were found in
XTE~J1550--564 (\pcite{2001ApJS..132..377H}). In general the energy
dependence and time lag behaviour of the QPOs in GX~5--1
(\pcite{1999A&A...343..197V}) is different from what was found in
XTE~J1550--564 (\pcite{wihova1999}). On the Flaring Branch, similar
to the High State in black hole candidates, the fractional rms
amplitude of the variability is low, and QPOs are weak or absent (in
GX~17+2 and Sco~X--1, just as in XTE~J1550--564 a QPO with a centroid
frequency of $\sim$18 Hz is found, but whether they are related is
unclear). The similarities between the Horizontal Branch variability
and that of the Intermediate/Low State have been pointed out before
(\pcite{va1994}; \pcite{1994ApJS...92..511V};
\pcite{wiva1999}). Furthermore, the trend of a decrease in
variability timescale as the source approaches the Flaring Branch/High
State is the same in black hole candidates as in GX~5--1 and
4U~1608--52 (cf. \pcite{2001ApJS..132..377H};
\pcite{mevava1998b}).
\newline
\noindent
To further investigate the similarities between a black hole candidate
(XTE~J1550--564), a Z source (GX~5--1), and an atoll source
(4U~1608--52) we created colour--colour diagrams (CDs) for the three
sources in two ways. First, we used the colour definitions commonly
used for atoll and Z sources (soft and hard colour were defined as the
logarithm of the 3.6--6.2 / 2.5--3.6 keV and 9.8--16.0 / 6.2--9.8 keV
count rate ratio, respectively). These CDs are shown in the {\it left
panels} of Fig.~\ref{z_as_BHC}, \ref{1550_as_BHC}, and
\ref{1608_as_BHC}. Second, we created CDs in a similar way as
\scite{vave1994}, an approach often used for black hole candidates
(\pcite{1997ApJ...488L.109B}; \pcite{2001ApJS..132..377H};
although see also \pcite{mikiki1991}). We defined the hard and soft
colour as the 9.7--16.0/ 2--6.4 keV and 6.4--9.7/ 2--6.4 keV count rate
ratio for observations 1--41, respectively and that of the 9.7--15.8 /
2--6.6 keV and 6.6--9.7 / 2--6.6 keV count rate ratio for observations
42--76, respectively (Fig.~\ref{z_as_BHC}, {\it right panel}). In case
of the {\it right panels} for Fig.~\ref{1550_as_BHC} and
\ref{1608_as_BHC} the hard and soft colour were defined as the
16.0--19.4 / 2.2--6.2 keV, and 6.5--15.7 / 2.2--6.2 keV count rate
ratio, respectively. Note that in the {\it right panels} the soft
colour is plotted vs. the hard colour, as the convention, introduced by
\cite{ostriker77} to plot hard colour vs. soft colour was abandoned in
some of the recent black hole work. In the {\it right panels} the
points have been connected by a line. Lines bridging different parts
of the diagrams do not reflect sudden jumps or changes in source
state, but are due to gaps introduced by observational windowing.
\newline
\noindent
A similar structure as for XTE~J1550--564 was found in the CD of
GX~5--1 (Figure ~\ref{z_as_BHC}, \ref{1550_as_BHC} {\it right
panel}). The Flaring Branch is now vertical and located at the left
side of the figure, similar to the High State in XTE~J1550--564. The
Normal Branch connects to the Flaring Branch and points towards the
upper right corner of the diagram (both the hard colour and the soft
colour increase as the source moves up the Normal Branch). The
Horizontal Branch is a continuation of the Normal Branch, with only a
slight bend at the Normal Branch--Horizontal Branch junction. (It is
interesting to note that there is also no vertex present in GX~17+2 at
the Horizontal-- Normal Branch junction at energies above 14.8 keV,
Figure 3C of \pcite{homanetal2001}). The 'Dipping' Flaring Branch is
parallel to the Normal Branch but directed towards the lower left
corner. The 'Dipping' Flaring Branch has been found in the Z sources
Cyg~X--2, GX~5--1 and GX~340+0 (\pcite{1996AA...311..197K};
\pcite{1994A&A...289..795K}; \pcite{2000ApJ...537..374J}).  The CD
of the atoll source 4U~1608--52 again shows a similar structure as
found for XTE~J1550--564 (Fig.~\ref{1550_as_BHC}, \ref{1608_as_BHC}
{\it right panels}). The island state is found in the top right corner
of the {\it right panel} of Fig.~\ref{1608_as_BHC}, whereas the lower
banana branch connects to the upper banana branch in the lower left
corner. The upper banana branch seems similar to the black hole
candidate High State and the Z source Flaring Branch.
\newline
\noindent
The CD of the black hole candidate XTE~J1550--564, plotted using
colours typically applied for neutron star sources is reminiscent of
that of an atoll source, although more structure is observed
(Fig.~\ref{1550_as_BHC} {\it left panel}). The source was first found
in the upper right part of the CD as the outburst started. The path
traced by the source as the outburst progressed is indicated with
arrows. For a complete description of the outburst of XTE~J1550--564
we refer to \scite{cuzhch1999}, \scite{somcre2000b}, and
\scite{2001ApJS..132..377H}.
\newline
\noindent
\noindent
The similarities between the colour--colour diagrams of the Z source
GX~5--1, the atoll source 4U~1608--52, and the black hole candidate
XTE~J1550--564 strengthen the idea that some of the spectral
properties of LMXBs originate in the accretion disc and do not depend
on the presence of a solid surface. However, our analysis also
indicates that differences between spectral properties of the neutron
star Z and atoll sources and the black hole candidates do exist (e.g.,
the {\it left panel} of Fig.~\ref{z_as_BHC}, \ref{1550_as_BHC}, and
\ref{1608_as_BHC}) but may not clearly show up in representations of
the CD commonly used for black--hole candidates. The explanation for
this could be that in the ``black--hole colour scheme'' only three
independent colours are defined whereas in the ``neutron star colour
scheme'' four colours are usually defined.

\noindent
So, although the CDs of a Z or atoll source and that of a black hole
candidate appear similar in shape when plotted using colours typical
for black hole studies, their appearance is both dissimilar and more
complicated when plotted using colours typical for neutron star
studies. Whether the complications can be ascribed to the unique
nature of the sources considered can only be checked by comparing more
sources but the conclusion that dissimilar CDs can appear more similar
than they really are in the ``black hole'' representation can already
be drawn. Conclusions regarding the spectral state or properties of
the source based on the colours commonly used in black hole candidates
should be regarded with caution since an other choice of colours may
lead to a qualitatively different CD.

\section*{Acknowledgments}
This work was supported in part by the Netherlands
Organization for Scientific Research (NWO). This research has made use
of data obtained through the High Energy Astrophysics Science Archive
Research Center Online Service, provided by the NASA/Goddard Space
Flight Center. This work was supported by NWO Spinoza grant 08-0 to
E.P.J.van den Heuvel. P.G.J. would like to thank Rob Fender, Kieran
O'Brien, and Marc Klein Wolt for various discussions. RW was supported
by NASA through Chandra Postdoctoral Fellowship grant under PF9--10010
awarded by CXC, which is operated by SAO for NASA under contract
NAS8--39073.

\end{document}